\documentclass[useAMS,usenatbib,letterpaper]{mn2e}
\usepackage{graphicx,amsmath,color,amssymb}
\usepackage[pdftitle={On the OVI Abundance in the Circumgalactic Medium of Low-Redshift Galaxies},pdfauthor={J. Suresh}]{hyperref}
\usepackage[all]{hypcap}
\usepackage{subfigure}

\newcommand{\adsurl}[1]{\href{#1}{ADS}}

\renewcommand{\thefootnote}{\fnsymbol{footnote}}
   
\newlength{\tinyFigurewidth}
\newlength{\narrowFigurewidth}
\newlength{\Figurewidth}
\newlength{\newFigurewidth}
\newlength{\subwideFigurewidth}
\newlength{\wideFigurewidth}
\newlength{\widerFigurewidth}
\newcommand{\eq}[1]
  {Eq.~(\ref{eqn:#1})}
\newcommand{\sect}[1]
  {Section~\ref{sec:#1}}

\newcommand{\fig}[1]
  {Figure~\ref{fig:#1}}

\newcommand{\Msun}{\, M_\odot}

\title[On the OVI Abundance in the CGM of Low-Redshift Galaxies]{On the OVI Abundance in the Circumgalactic Medium of Low-Redshift Galaxies}

\author[Joshua Suresh et al.]
{\parbox{\textwidth}{Joshua Suresh,$^{1}$\thanks{E-mail: \texttt{jsuresh@cfa.harvard.edu}}
Kate H. R. Rubin,$^{1}$
Rahul Kannan,$^{2}$
Jessica K. Werk,$^{3}$
Lars Hernquist,$^{1}$
Mark Vogelsberger$^{2}$}\vspace{0.4cm}\\
\parbox{\textwidth}{$^1$Harvard-Smithsonian Center for Astrophysics, 60 Garden Street, Cambridge, MA\\
$^{2}$Department of Physics, Kavli Institute for Astrophysics and Space Research, Massachusetts Institute of Technology, Cambridge, MA \\
$^{3}$UCO/Lick Observatory; University of California, Santa Cruz, CA, USA 
}}

\begin{document}

\pagenumbering{alph}
\date{}

\maketitle
\pagerange{\pageref{firstpage}--\pageref{lastpage}} \pubyear{2015}

\pagenumbering{arabic}
\label{firstpage}

\begin{abstract}

We analyze the mass, temperature, metal enrichment, and OVI abundance of the circumgalactic medium (CGM) around $z\sim 0.2$ galaxies of mass $10^9 M_\odot <M_\bigstar < 10^{11.5} M_\odot$ in the Illustris simulation.  Among star-forming galaxies, the mass, temperature, and metallicity of the CGM increase with stellar mass, driving an increase in the OVI column density profile of $\sim 0.5$ dex with each $0.5$ dex increase in stellar mass.  Observed OVI column density profiles exhibit a weaker mass dependence than predicted: the simulated OVI abundance profiles are consistent with those observed for star-forming galaxies of mass $M_\bigstar = 10^{10.5-11.5} M_\odot$, but underpredict the observed OVI abundances by $\gtrsim 0.8$ dex for lower-mass galaxies.  We suggest that this discrepancy may be alleviated with additional heating of the abundant cool gas in low-mass halos, or with increased numerical resolution capturing turbulent/conductive mixing layers between CGM phases.  Quenched galaxies of mass $M_\bigstar = 10^{10.5-11.5} M_\odot$ are found to have 0.3-0.8 dex lower OVI column density profiles than star-forming galaxies of the same mass, in qualitative agreement with the observed OVI abundance bimodality.  This offset is driven by AGN feedback, which quenches galaxies by heating the CGM and ejecting significant amounts of gas from the halo.  Finally, we find that the inclusion of the central galaxy's radiation field may enhance the photoionization of the CGM within $\sim 50$ kpc, further increasing the predicted OVI abundance around star-forming galaxies.

\end{abstract}

\begin{keywords}
circumgalactic medium -- intergalactic medium -- galaxies: formation -- methods: hydrodynamic simulations
\end{keywords}

\section{Introduction}
\renewcommand{\thefootnote}{\fnsymbol{footnote}}

O$^{5+}$, or OVI, is the most commonly-observed metal line absorber in
cosmological sightlines
\cite[e.g.][]{2004ApJ...606...92S,2008ApJ...679..194D,2008ApJS..177...39T,
  2009ApJS..182..378W}.  There are two important reasons for its
prominence.  First, after hydrogen and helium, oxygen is the third
most abundant element in the Universe.  Second, OVI produces a strong
absorption doublet redward of 1000 \AA ($\lambda$ 1031, 1037 \AA),
which greatly aids spectral identification.  OVI traces plasma in a
wide variety of environments, from ionized metals in the intergalactic
medium to high velocity cloud complexes within the Galactic halo.  The
highest column-density OVI absorbers ($N_\text{OVI} > 10^{13}$
cm$^{-2}$) are predicted to be associated with gas within a few
hundred kpc of galaxies
\citep{2009MNRAS.395.1875O,2011ApJ...731....6S}.

Using the Cosmic Origins Spectrograph on the Hubble Space Telescope,
\cite{2011Sci...334..948T} (henceforth T11) studied the abundance of
OVI around galaxies by carrying out a survey of QSO sightlines passing
close, in projection, to a sample of $z \sim 0.2$ galaxies.  The
selected galaxies span a wide range of stellar masses ($10^{9.5}
M_\odot \lesssim M_\bigstar \lesssim 10^{11.5} M_\odot$) and specific
star-formation rates (including red galaxies and star-forming galaxies
with specific star formation rates (sSFRs) up to $\sim 10^{-9}$
yr$^{-1}$).  T11 found that above their typical column density
detection limit of $N_\text{OVI} \gtrsim 10^{14}$ cm$^{-2}$, the 30
star-forming galaxies in the sample were mostly detected, but not so
for the majority of the 12 passive galaxies observed.  The velocity
centroids and widths of the OVI absorbers typically lie within the
inferred halo escape velocity, suggesting that the OVI arises from
bound gas within the virial radius.  Two important results follow.
First, star-forming galaxies have a roughly uniform OVI column density
profile with radius over two orders of magnitude in stellar mass
($10^{9.5} M_\odot \lesssim M_\bigstar \lesssim 10^{11.5} M_\odot$).
Second, the incidence of $N_\text{OVI} > 10^{14} $ cm$^{-2}$
absorption is significantly higher around star-forming galaxies than
passive galaxies in the stellar mass range $M_\bigstar \gtrsim
10^{10.5} M_\odot$.

To understand these two results, the origin of the OVI-tracing gas
must be understood.  The interpretation is complicated by the fact
that OVI typically features a significantly broader velocity profile
than nearby HI absorbers, making it difficult to constrain the total
mass or metallicity of the gas which is traced by OVI.  Furthermore,
the important ionization states OVII and OVIII may only be seen in
absorption in the X-ray range, which limits their detectability with
current instruments, thus making it difficult to constrain the
ionization state of the OVI-tracing gas. 

Various suggestions have been made for where the OVI ``phase"
originates.  First, it is possible that OVI traces material shocked
and entrained by supernova ejecta, which cools adiabatically as it
propagates from the central galaxy \citep[e.g.][]{Thompson:2015wy}.
Second, OVI may trace the shock-heated/virialized warm-hot
intergalactic medium (WHIM)
\citep{1999ApJ...519L.109C,2001ApJ...552..473D,2012ApJ...752...65N}.
Third, OVI may be the result of hot halo gas cooling through a thermal
instability \citep{2002ApJ...577..691H,2004MNRAS.355..694M}.  Fourth,
OVI may trace the interfaces between cold clouds and the hot
virialized halo, either in a turbulent mixing layer
\citep{1990MNRAS.244P..26B} or a conduction layer
\citep{2003ApJS..146..165S}.  A fifth alternative is that the OVI is
actually cooler gas ($T \sim 10^4$ K) that is photoionized by the
local galactic radiation field \citep[][]{2008ApJS..177...39T}.

The abundance of OVI around galaxies has been examined in a number of
studies which focused on cosmological zoom-in simulations of one or
two halos.  Within this context, the galactic feedback can be tuned to
successfully match the observed OVI profile, with each study finding
that their strongest feedback model was the most successful
\citep[e.g.,][]{2012MNRAS.425.1270S}.  \cite{2013MNRAS.430.1548H} came
to a similar conclusion upon examination of the OVI profile around a
single simulated MW-like halo, ultimately inferring that it was
impossible to reproduce the observed OVI profile without modifying the
temperature of the CGM by hand.  \cite{2015arXiv150707002L} studied
the ion abundances around a single MW-mass halo with various feedback
prescriptions, finding that the feedback required to match the
observed OVI profile was sufficiently strong that it made the stellar
component of the galaxy unrealistically large and spheroid-dominated.
Since these studies each examined only one or two halos, they have not
shown that these feedback prescriptions successfully generate a
galactic population which matches observational constraints when
applied to a full cosmological sample.  Moreover, all these
investigations neglected AGN feedback.

\cite{2015arXiv150302084F} carried out a larger, statistical
comparison with a cosmological hydrodynamic simulation, similar in
spirit to our work.  They found that their model underpredicts the
amount of OVI in the CGM.  However, it is unclear whether the
discrepancy arises universally for all galaxies in the simulation or
only for galaxies in a small mass range.  Furthermore, they found no
significant difference in the OVI absorption around passive and
star-forming galaxies within the simulation.  This may be due to their
use of an ``artificial" quenching mechanism which has negligible
impact on the circumgalactic gas, unlike an explicit AGN feedback
model \citep[e.g.,][]{2015MNRAS.448L..30C}.  The simulations of
\cite{2015arXiv150302084F} may also suffer from numerical issues with
their adopted hydrodynamic solver
\citep[e.g.,][]{2012MNRAS.424.2999S,2012MNRAS.425.2027K,2013MNRAS.429.3353N,2015ApJ...800....6Z}.

In this paper, we use the Illustris simulation
\citep{2014Natur.509..177V, 2014MNRAS.444.1518V, 2015arXiv150400362N}
to explore the relationship between galaxies and the CGM, with the
goal of understanding the T11 results.  Illustris has a subgrid model
for processes thought to be relevant to galaxy formation, including
galactic winds driven by supernovae as well as AGN feedback.  The
simulation contains a large statistical sample of galaxies having a
wide range of environments, histories, and star-formation rates, and
shows reasonable agreement with observed $z \sim 0$ galactic scaling
relations such as the galaxy mass function and the star-formation main
sequence.  In particular, since the galactic wind model of Illustris
has a velocity scaling with halo mass such that the majority of wind
material is retained within the halo \citep{2015MNRAS.448..895S}, the
simulation offers a ``maximal" scenario for metal enrichment of small
galaxies.  Conversely, for more massive galaxies, the powerful AGN
feedback model of Illustris offers the opportunity to test whether the
same mechanism that quenches massive galaxies also has a significant
impact on the CGM.

\section{Methods}
\label{sec:methods}

\subsection{Simulation Details}

The Illustris simulation \citep{2014MNRAS.444.1518V} is a large
cosmological hydrodynamic simulation run with the moving-mesh
\textsc{arepo} code \citep{Springel:2010hx}, using the galaxy
formation model presented in \cite{2013MNRAS.436.3031V} (henceforth
V13).  \textsc{arepo} employs a finite-volume scheme over a grid of
cells defined using a Voronoi tessellation of space.  The
mesh-generating points which are employed to construct the
tessellation move approximately with the bulk motion of the fluid.
This allows the code to benefit from the strengths of both Eulerian
and Lagrangian approaches
\citep{Springel:2010hx,2012MNRAS.425.3024V,2012MNRAS.427.2224T}.

The size of the periodic simulation volume in Illustris is 75 $h^{-1}$
Mpc, and has $1820^3$ dark matter particles and $1820^3$ initial gas
resolution elements.  This latter number can vary over time as the
code adaptively refines or de-refines gas cells, in order to keep a
roughly fixed mass resolution.  The dark matter mass resolution is
$6.26 \times 10^6 M_\odot$ and the initial gas mass resolution is
$1.26 \times 10^6 M_\odot$.  The Illustris simulation adopts the
WMAP-7 cosmology ($\Omega_{\Lambda,0} = 0.73$, $\Omega_{m,0} = 0.27$,
$\Omega_{b,0} = 0.0456$, $\sigma_8 = 0.81$ and $h = 0.704$).
Illustris provides a good match to the low-redshift galaxy population,
especially the $z \sim 0$ stellar mass function, star formation main
sequence, and Hubble sequence
\citep{2014MNRAS.444.1518V,2014MNRAS.445..175G,2015MNRAS.447.3548S}.
In addition, the Illustris physics model has been shown to give good
agreement with damped Lyman-$\alpha$ absorber abundances,
metallicities, and kinematics at higher redshifts
\citep{2014MNRAS.445.2313B,2015MNRAS.447.1834B}.

For details of the numerical and feedback parameters used in this
paper, we refer the reader to V13.  Here we briefly review some
salient features.  Primordial and metal line cooling are included in
the presence of a spatially uniform, time-varying UV background
\citep{Katz:1996tu,2009MNRAS.399..574W,2009ApJ...703.1416F},
accounting for gas self-shielding following
\cite{2013MNRAS.430.2427R}.  Star particles are formed using a
two-phase description of the interstellar medium (ISM)
\citep{2003MNRAS.339..289S}.  Stellar evolution and chemical
enrichment are followed as described in V13, explicitly tracking the
following 9 elements: H, He, C, N, O, Ne, Mg, Si, Fe.  The cooling
prescription assumes ionization equilibrium, which is reasonable at
low redshifts, by which time any AGN-induced non-equilibrium ``fossil
zones" have likely re-equilibrated \citep{2013MNRAS.434.1063O}.

Winds are launched from star-forming regions, with the wind speed set
at 3.7 times the 1D local halo velocity dispersion.  This choice was
made to ensure that the wind material is largely recycled, maintaining
a sufficient cosmic star formation rate at $z \lesssim 1$ (see V13).
Another, related consequence of this choice is that the winds
typically do not escape the halo \citep{2015MNRAS.448..895S}.  The
mass loading (ratio of the wind mass flux to the SFR) is $\eta \propto
\sigma_\text{DM,1D}^{-2}$, where $\sigma_\text{DM,1D}$ is the local
one-dimensional DM velocity dispersion.

In order to ensure that the winds escape the dense ISM, wind cells are
hydrodynamically decoupled and allowed to propagate as particles until
they either travel for a set time, or reach sufficiently low-density
regions (the winds typically recouple hydrodynamically within a few
kpc of the galaxy).  The wind particle mass, momentum, energy, and
metals are then deposited into the nearest gas cell.  The winds are
launched cold; however we have found that the wind particles
immediately shock on the ambient gas and thermalize, re-distributing
their energy between kinetic and thermal components.  In other words,
a cold fast wind and a hot slow wind with the same overall energy per
unit mass will have a similar impact on the phase structure of the
CGM; what matters more is the overall energy per unit mass of the wind
\citep{2015MNRAS.448..895S}.

Each wind particle is assigned a metallicity equal to some fraction of
that of the ambient ISM from which it was launched.  The fiducial wind
metal-loading factor of 0.4 was chosen to allow the simulations to
match the galaxy mass-metallicity relation.  We note that the winds
are effective enough that the CGM is enriched at a level comparable to
observational estimates (see \sect{SFMS}).

The Illustris physics model also includes AGN feedback from
supermassive black holes. Black hole (BH) particles are seeded at
early times and allowed to grow through accretion and merging events.
Since the scale of the event horizon is always well below the
resolution of the simulation, a sub-grid feedback model is employed
which is based on methods developed in earlier studies
\citep{2005Natur.433..604D,2008ApJ...676...33D,2005MNRAS.361..776S,2007MNRAS.380..877S}.
BH feedback has two modes, depending on the accretion rate.  At high
accretion rates, the BH enters a ``quasar-mode", in which a fraction
of the accretion energy is thermally coupled to the nearby gas.  At
lower accretion rates, the BH enters ``radio-mode" feedback, which
mimics observations of radio jets injecting energy on larger scales,
by inflating hot bubbles which are hydrodynamically buoyant.  In our
implementation, this is carried out by injecting energy in
randomly-distributed bubbles at a distance offset of tens of kpc away
from the center of the galaxy.  Finally, we approximate the effect of
AGN irradiation by modifying the heating and cooling rates for gas
cells near the BH.

The time an average BH spends in the quasar mode is short compared to
the time spent in the low-accretion radio mode
\citep{2015MNRAS.452..575S}.  V13 found that the radio mode feedback
has, by far, the most significant impact on the $z < 1$ cosmic
star-formation rate compared to the other elements of the AGN physics
model.  This is due to the AGN feedback becoming increasingly more
important in large halos ($M_\text{halo} \gtrsim 10^{12} M_\odot$),
playing a major role in the quenching of massive galaxies at late
times.

Since our work deals primarily with the metal enrichment of the
circumgalactic medium, we briefly review the relevant subgrid model in
Illustris.  Each star particle which forms in our simulation
corresponds to a single-age stellar population (SSP), which produces
metals over time depending on its IMF and age (for example, an SSP
with a top-heavy IMF will deposit its metals at a higher rate as the
more massive stars evolve faster).  We assume a Chabrier IMF.
Enrichment from AGB stellar mass loss, core-collapse supernovae, and
Type Ia supernovae are included, as these are the primary channels by
which a stellar population's metal production enriches the nearby ISM
\citep[assumed yields are discussed in][]{2013MNRAS.436.3031V}.  As
metals are generated over time by a star particle, they are deposited
into nearby gas cells, after which the metals are treated as a passive
tracer which is hydrodynamically advected with the gas.  This mixing
scheme is a strength of the code in the context of CGM studies and is
not shared by most smoothed-particle hydrodynamic implementations
\citep{2012MNRAS.420..829O}.  Feedback processes thus transport metals
into the CGM by ejecting enriched gas from the galaxy into the halo.

\subsection{Identifying and Selecting Galaxies}

Dark matter halos are identified using the \textsc{subfind} algorithm
\citep{2001MNRAS.328..726S}, with a friends-of-friends (FoF)
group-finding approach, adopting a linking length of 0.2 times the
mean interparticle separation.  In this paper we limit our analysis to
``central galaxies", or the galaxies which inhabit the largest halo of
each FoF group.  This allows us to incorporate the effects of any
satellite galaxies on the shared CGM (e.g. small satellites helping to
enrich the CGM through stellar outflows or tidal stripping).  We
select all central galaxies with stellar mass $M_\bigstar> 10^{9}
\Msun$ within twice the half-mass radius, which are resolved galaxies
with $>800$ star particles.  We define the virial radius of each
central galaxy as the radius within which the average density exceeds
200 times the critical density (we will refer to this radius as
$R_{200}$).  When computing global CGM properties, we define the CGM
as all gas which is not star-forming and which is outside of twice the
half-mass radius of any galaxy (central or satellite) in the halo.
Conversely, when computing OVI maps around the halo, we include all
gas, including the galaxies' ISM.

\begin{figure}
\label{fig:galmass}
\includegraphics[width=\newFigurewidth]{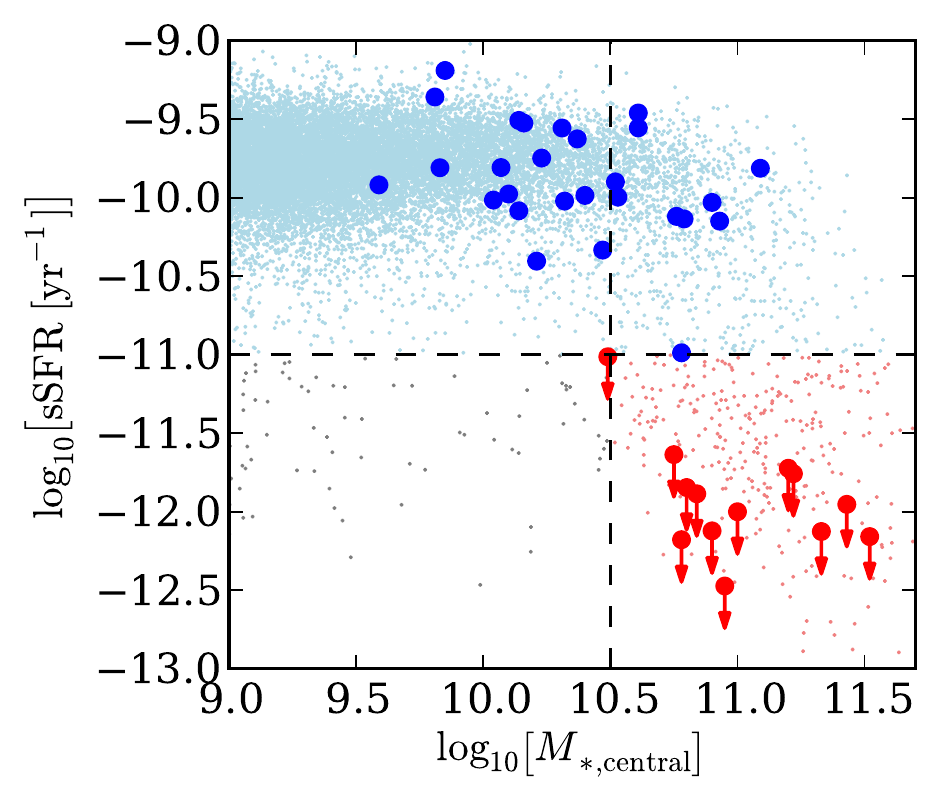}
\caption{Comparison between the central galaxies in the Illustris
simulation (small points) and the galaxies selected in the COS-Halos
survey (large points).  The vertical line shows
$M_\bigstar=10^{10.5}$, and the horizontal line shows the COS-Halos
threshold between star-forming and passive galaxies, namely
sSFR$=10^{-11}$ yr$^{-1}$.  The gray points in the bottom left
quadrant are passive galaxies which are below the minimum passive
galaxy mass in the COS-Halos survey; these are excluded from our
analysis.  Note that the additional galaxies from
\protect\cite{2015MNRAS.449.3263J}} are not shown, as they do not
have have measured star-formation rates, only the distinction of
``early-type" or ``late-type" based on emission line diagnostics.
\end{figure}

\cite{2011Sci...334..948T} selected a galaxy sample with a median
redshift $\left< z \right>\sim 0.2$ close to known quasar sightlines.
The galaxies were chosen to sample a wide range of stellar masses
($M_\bigstar > 10^{9.5} M_\odot$) and specific star formation rates.
In \fig{galmass} we compare the stellar masses and specific
star-formation rates of the Illustris central-galaxy population and
the COS-Halos galaxy sample.  In the few cases where their follow-up
galactic spectroscopy revealed a more massive galaxy at the same
redshift, they changed their adopted impact parameter to reflect that
of the more massive galaxy.  Specifically, they select the most
massive galaxy within 300 kpc transverse, and $|500 \text{km/s}|$
(which corresponds to about $\pm 7 $Mpc along the line of sight).
This is similar but entirely equivalent to our approach of considering
only central galaxies of the FoF groups, and treating the satellite
contributions as part of the CGM.  We test whether our method differs
significantly from T11's by choosing 10000 random positions (analogous
to quasar sightlines) within our simulation volume, and checking
whether the most massive galaxy and the closest galaxy are the same
within the observational ``cylinder'' (radius 300 kpc, length 14 Mpc).
For galaxies with $M_\bigstar > 10^{9} M_\odot$, the closest galaxy
and the most massive galaxy in the cylinder are mismatched only $\sim
5\%$ of the time.  Hence, our central galaxy-focused approach is
largely equivalent to the observational galaxy-selection technique.

\begin{figure*}
\includegraphics[width=\wideFigurewidth]{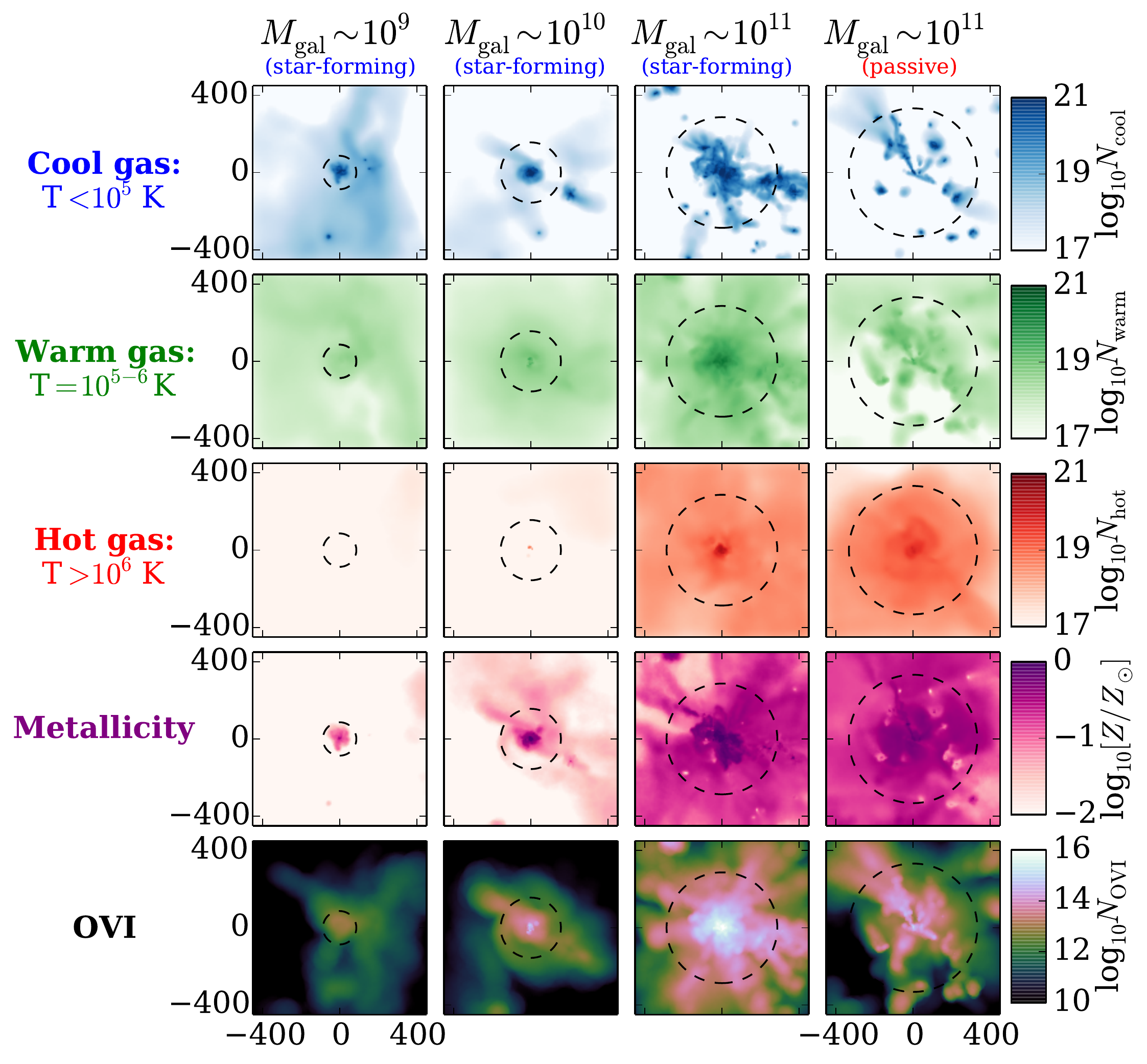}
\label{fig:images}
\caption{Distribution, metallicity, and OVI abundance of gas around
four representative galaxies from the simulation volume at $z =0.2$.
Each column of images corresponds to a single galaxy, with the
properties shown at the top of the column.  The side length is
roughly $\pm 400$ proper kpc, centered on the galaxy; the dashed
circles show the virial radius.  The CGM has multiple phases at all
galaxy masses, with cool clumpy gas embedded in a warmer ambient
medium.  The total mass of both cool and hot gas increases with
galaxy mass, as does the metallicity of the gas.  This translates to
higher OVI column densities for more massive galaxies (shown in
bottom row).  The passive galaxy shows slightly decreased OVI, which
is due mainly to the diminished warm gas content in the halo
outskirts compared to the star-forming galaxy of the same mass.
Conversely, the passive halo has more hot gas, since it possesses a
larger central AGN and a more massive halo.}
\end{figure*}

We also include relevant measurements from \cite{2015MNRAS.449.3263J}.
These authors compiled an observational dataset from their own
absorption-blind survey, combined with quasar sightlines around 11
massive galaxies in the Sloan Digital Sky Survey
\citep{2000AJ....120.1579Y}, as well as the COS-Halos sample.  Since
their focus was to study the role of environment in circumgalactic OVI
and HI abundances, many of the galaxies in their sample are identified
as being ``non-isolated", which they define as having a spectroscopic
neighbor within a projected distance of less than 500 kpc, a radial
velocity difference of $\left| \Delta v \right| < 300$ km s$^{-1}$,
and with stellar mass of at least one-third of that of the survey
galaxy.  We include only isolated galaxies from
\cite{2015MNRAS.449.3263J} in our comparisons.

\cite{2015MNRAS.449.3263J} do not explicitly derive SFRs as in the
COS-Halos survey, but instead identify galaxies as ``late-type" or
``early-type" derived from absorption/emission line diagnostics, based
on methods developed in \cite{2009ApJ...701.1219C}.  We include the
``late-type" (``early-type") points along with the star-forming
(passive) galaxies of T11, and the two datasets typically exhibit
similar trends.  Finally, we note that apart from the COS-Halos data,
which are at impact parameters of $< 150$ kpc, most of the sightlines
of \cite{2015MNRAS.449.3263J} are at much larger distances, out to
$\sim 1$ Mpc.  However, since virtually all of the sightlines at large
distances are non-detections in OVI for both star-forming and passive
galaxies, we focus only on sightlines within 300 kpc when comparing to
the the simulation.

Finally, we include also measurements from \cite{2011ApJ...740...91P},
who carried out a survey of QSO sightlines around galaxies of
luminosity $L > 0.1 L_*$ with known redshifts.  As for the other
datasets, we exclude non-isolated galaxies by selecting only the
brightest galaxy within 200 kpc and $|\delta v| < 400$ km s$^{-1}$, as
given in their paper.  Unfortunately, they did not have sufficient
photometry to determine galaxy masses, so the galaxies are tagged by
their luminosity and by being ``early-type'' or ``late-type'' based on
absorption/emission line diagnostics.  Not knowing the precise masses
of the galaxies in this sample makes detailed interpretation
difficult, but when possible we include them with the other
observations.

\subsection{Computing OVI Abundances}
\label{sec:cloudy}

The total oxygen abundance in each gas cell is a direct output of the
simulation.  We calculate the ionization fraction of the oxygen using
Cloudy 13.03 \citep{2013RMxAA..49..137F}, with the extragalactic
ultraviolet background spectral energy distribution (UVB SED)
\citep[for which we adopt][]{2009ApJ...703.1416F} used as an input to
the code, assuming ionization equilibrium.  Both photo-ionization and
collisional ionization are accounted for.  All of the results in our
paper implement only the extragalactic UVB, except for \sect{SED}
where we explore the effect of adding a local galaxy SED on top of the
extragalactic background.  We run Cloudy in single-zone mode, which
treats the gas as a single slab illuminated on one side by the given
radiation field.  We use slightly modified versions of Simeon Bird's
Cloudy scripts, which are freely available at
\url{https://github.com/sbird/cloudy_tables}.

With the OVI content of each gas cell in hand, we compute radial OVI
profiles in order to compare to observations.  We project the cells
onto a uniform grid around each halo of extending $400$ transverse
proper kpc.  Interpolation of the gas density onto the grid is
performed using an SPH kernel with size corresponding to the gas cell
size as in \cite{2013MNRAS.429.3341B}.  We treat each grid cell as a
single sight line, analogous to the random locations of quasar
sightlines around physically unrelated galaxies
\citep[e.g.,][]{1996ApJ...457L..57K}.

We include all gas cells within a line-of-sight velocity window around
the galaxy position, with the velocity width chosen to roughly
correspond to that used by the observational experiments.  In the
COS-Halos sample, the velocity centroid of the observed OVI absorbers
is typically within $\pm 150$ km/s of the galactic redshift.  This
velocity window corresponds to a large line-of-sight distance in a
pure Hubble flow ($\sim \pm 2.7$ Mpc), so it is not clear a priori
that the gas must be located within the halo simply by velocity
arguments.  Nevertheless, we have confirmed that, in the simulation,
even when including all gas within this large line-of-sight distance,
the vast majority of the OVI absorption occurs within the virial
radius of the halo.  The only exception is at impact parameters $b
\gtrsim 200$ kpc (which exceeds the maximum impact parameter of 150
kpc probed by T11), where interloper absorbers at large distances from
the halo can boost the inferred OVI absorption, especially around
small galaxies.  We adopt the velocity window $\pm 150$ km/s when
computing all OVI radial profiles.  \fig{images} shows gas properties
around four representative galaxies in the simulation, specifically
the total column density in three temperature bins, the metallicity,
and the total OVI abundance.  Once we have computed the grid of OVI
column densities for each galaxy, we compile all grids for a given
galaxy population.  We then compute a radial profile for the
population by taking the median column density of all the sightlines
in radial bins of 20 kpc.

\section{Results}
\label{sec:results}

The mass $M_{X,j}$ of the $j$th ionization state of element $X$ in a gas cell is given by:
\begin{equation}
	M_{X,j} = M_\text{gas} \left(\frac{M_\text{metals}}{M_\text{gas}}\right) \left(\frac{M_\text{X}}{M_\text{metals}}\right) \left(\frac{M_\text{X,j}}{M_\text{X}}\right)
	\label{eqn:master}
\end{equation}
This means that the total CGM OVI abundance depends on: 1) CGM mass,
2) CGM metallicity, 3) oxygen-to-metal ratio, and 4) OVI ionization
fraction.  We now explore how each of the above factors varies with
stellar mass and star-formation rate in order to determine the origin
of the observed dependence of OVI abundance on these galaxy
properties.

\subsection{The Star Formation Main Sequence: $10^9 M_\odot < M_\bigstar < 10^{11.5} M_\odot$} 
\label{sec:SFMS}

We begin by focusing on the star-forming galaxies at $z=0.2$ over a
wide range in galaxy stellar mass: $10^9 M_\odot < M_\bigstar <
10^{11.5} M_\odot$.  Here we define ``star-forming" by a simple cut in
the specific star formation rate: sSFR$ > 10^{-11}$ yr$^{-1}$, which
is the same threshold used in the COS-Halos survey.  None of the
results in this section include passive galaxies (sSFR$ < 10^{-11}$
yr$^{-1}$).

\subsubsection{Global CGM properties}
\label{sec:SFMS_global}

In this subsection, we define the CGM as all gas within $R_\text{vir}$
that is not star-forming and that is outside of twice the half-mass
radius of any galaxy (central or satellite) in the halo.

\begin{figure}
\includegraphics[width=\newFigurewidth]{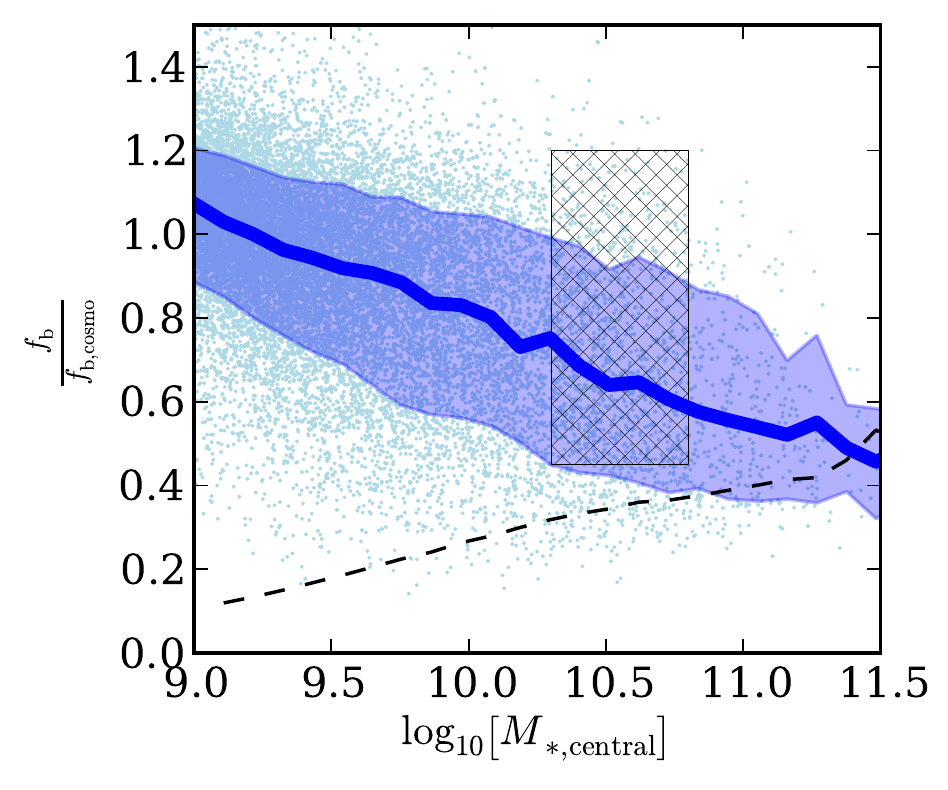}
\label{fig:baryfrac_SFMS}
\caption{Baryon fraction within $R_{200}$ as a function of galaxy
stellar mass.  Individual galaxies are plotted, while the solid blue
curve and blue shaded region indicate the median and $\pm 1 \sigma$
values of the distribution, respectively.  The hatched region
denotes the estimated baryon fraction for star-forming galaxies with
$M_\text{halo} \sim 10^{12.2} M_\odot$ from
\protect\cite{2014ApJ...792....8W}.  The dashed curve shows the same
relation from \protect\cite{2015MNRAS.451.1247S}.  Smaller galaxies
have retained virtually all of their baryons, while more massive
galaxies have kept about 50-60\% (due to mass loss from AGN
activity).  The discrepancy between our model and that of
\protect\cite{2015MNRAS.451.1247S} highlights the fact that the
baryon fraction of star-forming galaxies is a feedback-dependent
quantity.  The star-forming galaxies in the simulation retain the
majority of their baryons since the fiducial Illustris wind model
does not efficiently eject baryons from halos.}
\end{figure}

\fig{baryfrac_SFMS} shows the baryon fraction as a ratio of the cosmic
baryon fraction within $R_{200}$ versus galaxy stellar mass.  (Note
that baryon fractions can exceed unity since gas is dissipative while
dark matter is collisionless.)  Our wind model is tuned such that the
winds typically do not escape the halo \citep{2015MNRAS.448..895S},
which means that in small halos, where AGN feedback is not
significant, the halo baryon fraction is close to unity.  Conversely
for higher mass galaxies, the AGN can eject baryons out of the halo
through radio-mode feedback over time
\citep{2014MNRAS.445..175G,2015MNRAS.448..895S}.  Hence more massive
galaxies tend to have lower baryon fractions, with values around
$50\%$ in halos which host star-forming $M_\bigstar \sim 10^{11}
M_\odot$ galaxies.  This range in baryon fractions is consistent with
the findings of \cite{2014ApJ...792....8W}, who found that
star-forming galaxies in halos of mass $M_\text{halo}\sim10^{12.2}
M_\odot$ (and galaxy mass of $M_\bigstar \sim 10^{10.6}$) have
retained somewhere between $45-120\%$ of their baryons; this range is
denoted by the hatched region in \fig{baryfrac_SFMS}.  We note that
this agreement is not a trivial result, and is feedback-dependent.
For example, the dashed curve in \fig{baryfrac_SFMS} shows the baryon
fraction derived from the EAGLE simulation, as published in
\cite{2015MNRAS.451.1247S}, where we have converted from halo mass to
stellar mass using the relation given in \cite{2013ApJ...770...57B}.
\cite{2015MNRAS.451.1247S} find the opposite relationship, where
smaller galaxies have lower baryon fractions than larger galaxies.  We
discuss the implications of this difference in \sect{discussion}.

\begin{figure}
\includegraphics[width=\newFigurewidth]{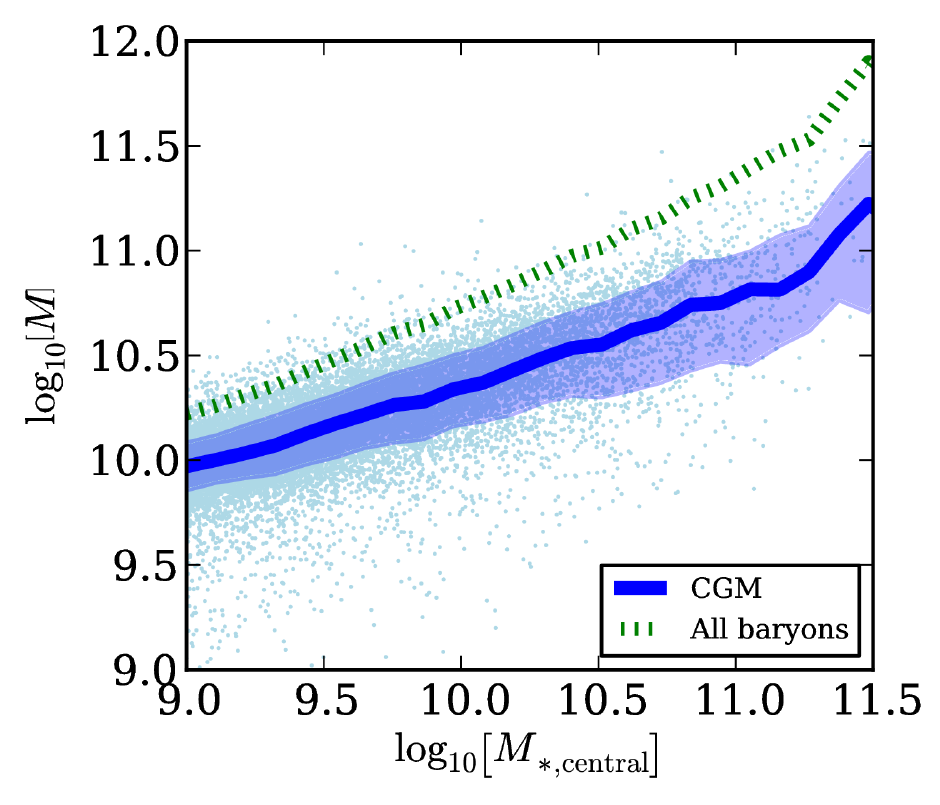}
\label{fig:CGM_mass_SFMS}
\caption{CGM mass within $R_{200}$ as a function of galaxy stellar
mass.  The solid line shows the binned median.  The blue curve and
shaded region show the median and $\pm 1 \sigma$ values of the CGM 
mass distribution, and the dashed green curve shows the median of 
the total baryonic mass within $R_{200}$.}
\end{figure}

\fig{CGM_mass_SFMS} shows the corresponding CGM (i.e. non-ISM) mass
within $R_{200}$ as a function of galaxy stellar mass.  The median
curve is well-approximated by the following linear fit:
$\log_{10}\left[M_\text{CGM} \right] = 0.44 \log_{10}\left[M_\bigstar
  \right] + 6.0$.  The CGM is a significant reservoir of baryons for
all galaxy masses, which is unsurprising, given that the halos of
star-forming galaxies retain their baryons to within a factor of $\sim
2$ (see \fig{baryfrac_SFMS}).  The ``break-even point" where the CGM
mass is comparable to the stellar mass of the galaxy occurs for
galaxies with $M_\bigstar \sim 10^{10.7} M_\odot$.  Below (above) this
galaxy mass, the $M_\text{CGM}$ is greater than (less than)
$M_\bigstar$.

\begin{figure}
\includegraphics[width=\newFigurewidth]{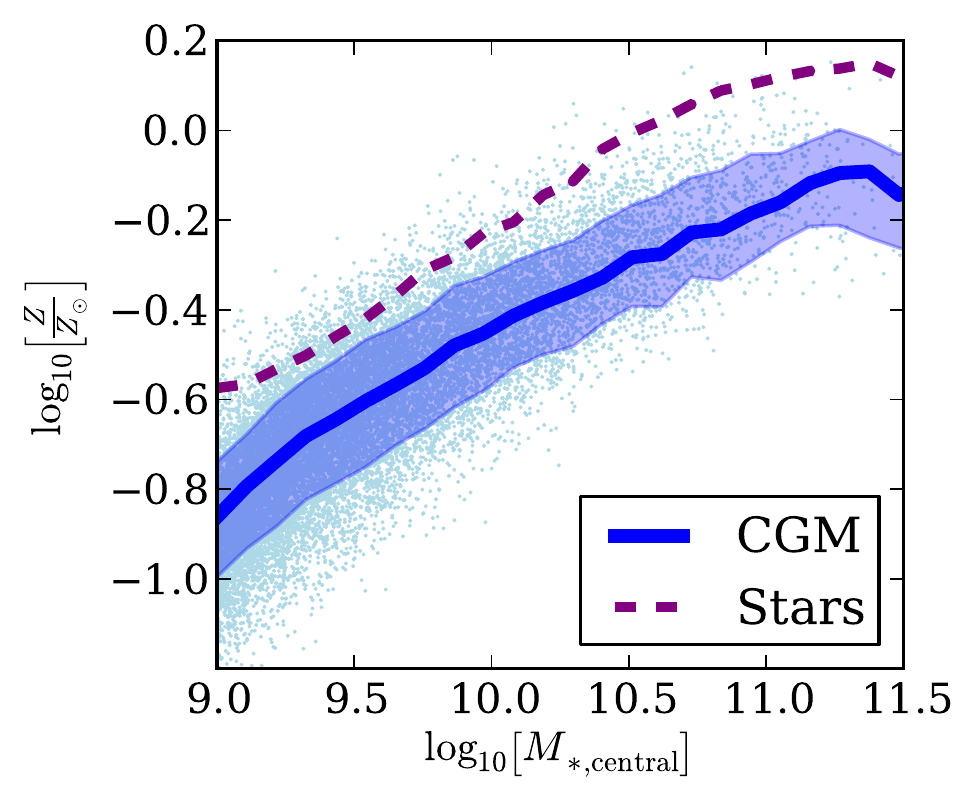}
\label{fig:z_CGM_SFMS}
\caption{CGM metallicity within $R_{200}$ as a function of galaxy
stellar mass.  The blue solid line and shaded region show the
binned median and $\pm 1 \sigma$ values for the CGM.  The dashed
line shows the median galaxy stellar metallicity for the central
galaxy.}
\end{figure}

\fig{z_CGM_SFMS} shows the CGM metallicity within $R_{200}$ as a
function of galaxy stellar mass.  The CGM metallicity is clearly
``aware" of the galaxy mass-metallicity relationship.  We find that
the CGM metallicity mirrors the galaxy metallicity: over 2.5 decades
in galaxy mass, the median CGM metallicity lies below the median
galaxy metallicity at a constant offset of $\sim 0.2-0.3$ dex.

The total metal content of the CGM and galaxies agree well with the
estimates of \cite{2013AAS...22132606P}, who compiled observational
measurements of the total metal content in the low-ion, OVI-traced,
X-ray, and dust components of the CGM.  \cite{2013AAS...22132606P}
found that for star-forming galaxies within the stellar mass range
$M_\bigstar=10^{9.3-10.8} M_\odot$, the total CGM metal mass is
$M_\text{Z,CGM} = 10^{7.7-8.5}$.  Over the same range in stellar mass,
the simulation gives a CGM metal mass of $M_\text{Z,CGM} =
10^{7.6-8.8}$.

\begin{figure}
\includegraphics[width=\newFigurewidth]{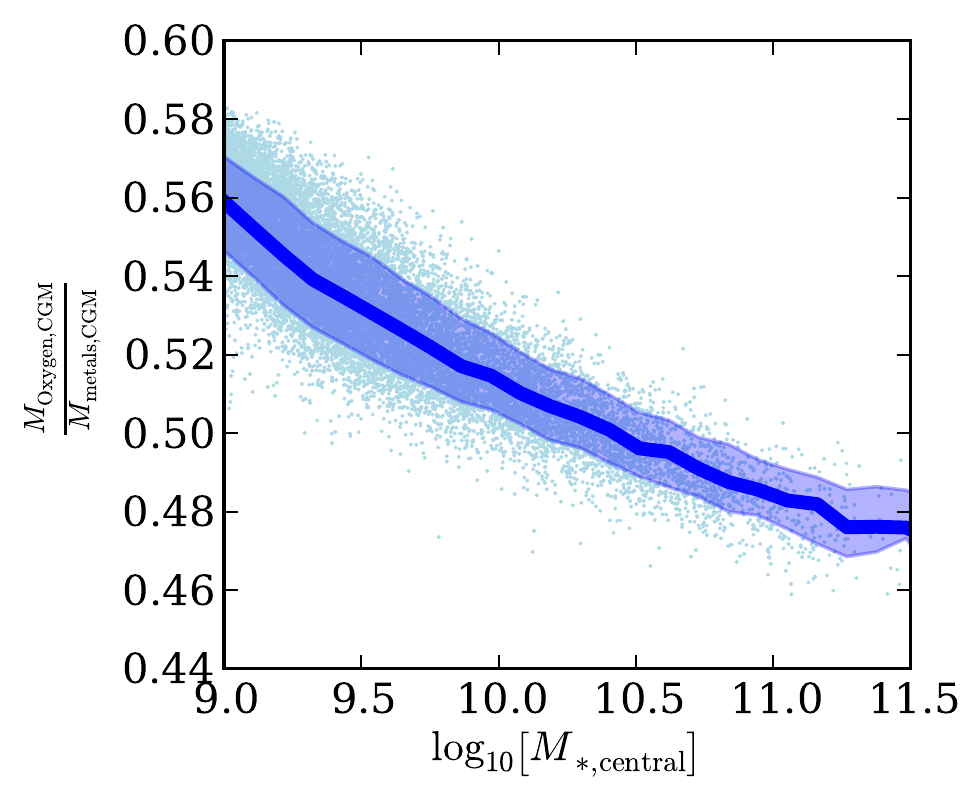}
\label{fig:O_metal_ratio}
\caption{Ratio of oxygen mass to metal mass within the CGM, as a
function of galaxy stellar mass.  Smaller galaxies have a slight
alpha-enhancement over massive galaxies, as the latter have had more
time for Type Ia supernovae to release iron into the CGM.  This
enhancement of oxygen is a much weaker trend with galaxy mass than
CGM mass, metallicity, or OVI ionization fraction.}
\end{figure}

At fixed metallicity, the relative abundance of individual elements
can vary based on, for example, the relative numbers of core-collapse
and Type Ia supernovae that have occurred.  In \fig{O_metal_ratio}, we
show the mass-weighted oxygen-to-metal ratio within $R_{200}$.  There
is a monotonic decrease in the amount of oxygen per unit metal with
increasing stellar mass.  This trend is driven by the fact that larger
galaxies have had more time for Type Ia supernovae to contribute
significant amounts of iron to the CGM, which lowers the alpha
enhancement.  Comparing to Figures \ref{fig:CGM_mass_SFMS},
\ref{fig:z_CGM_SFMS}, and \ref{fig:O_metal_ratio}, we find that
the variation in the alpha-enhancement of the CGM is a small effect
compared to the trends of CGM mass, metallicity, and ionization state
with galaxy stellar mass.

\begin{figure}
\label{fig:T_bins_SFMS}
\includegraphics[width=\newFigurewidth]{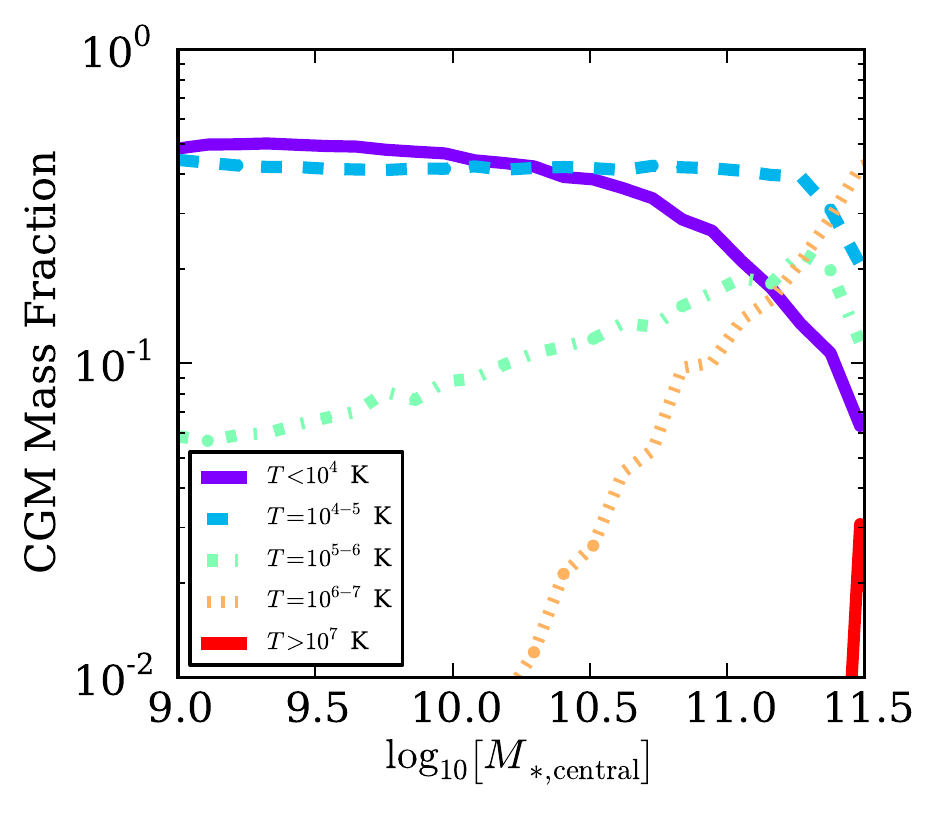}
\caption{Fraction of all CGM gas in different temperature ranges.  The
CGM is multiphase in nature at all galaxy masses, with a significant
fraction of gas with $T < 10^{5}$ K in both small and large halos.
Higher mass galaxies inhabit more massive halos, which heat some of 
the cold gas to $T=10^{5-6}$, thus boosting the OVI abundance, since
the OVI-tracing gas is predominantly in this temperature bin (see
\protect\fig{O6_mass_T_bins_SFMS}).  At the highest galaxy masses,
the fraction of gas within the range $T=10^{5-6}$ begins to drop 
as the fraction of hot gas ($T > 10^{6}$ K) rises.}
\end{figure}

In \fig{T_bins_SFMS}, we show the mass fractions of CGM gas in
different temperature ranges.  This figure highlights that the CGM has
a multiphase character at all galaxy masses
\citep[e.g.,][]{2015arXiv150302665N}.  Larger galaxies inhabit more
massive halos with deeper gravitational potential wells, so that they
have a higher fraction of hotter gas.  Nevertheless, there is a
substantial fraction of cool ($T < 10^{5} K$), non-ISM gas within
$R_{200}$, from $\sim 70\%$ of all gas in $M_\bigstar \sim 10^9
M_\odot$ galaxy halos to $\sim 25\%$ in $M_\bigstar \sim 10^{11.5}
M_\odot$ galaxy halos.  This gas likely originates from ongoing
accretion, galactic fountains, and from cooling instabilities in the
warm gas due to the upturn of the cooling function at $\sim 10^5$ K.
We find that the metallicity of the warmer gas is typically lower than
that of the cool gas, but that the metallicity of these phases
converges at higher galaxy mass.  For example, the metallicity of the
$T=10^{5-6}$ K gas is about 0.7 dex below that of the $T < 10^5$ K gas
for $M_\bigstar \sim 10^{9} M_\odot$ galaxies, 0.4 dex below at
$M_\bigstar \sim 10^{10} M_\odot$, and 0.05 dex below at $M_\bigstar
\sim 10^{11} M_\odot$.

\begin{figure}
\includegraphics[width=\Figurewidth]{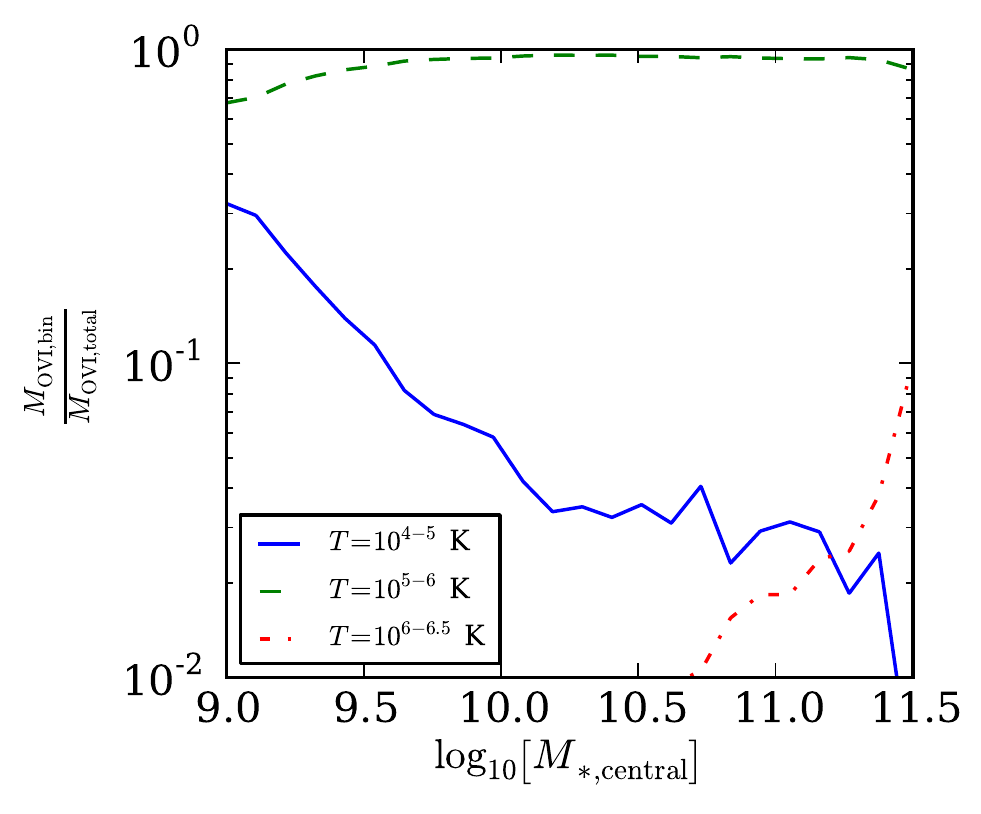}
\label{fig:O6_mass_T_bins_SFMS}
\caption{Fraction of total circumgalactic OVI which is in each
temperature range.  At all galaxy masses, the circumgalactic OVI 
is dominated by gas with temperatures $T=10^{5-6}$ K, implying
collisional ionization.}
\end{figure}

However, OVI does not trace the full range of circumgalactic gas
temperatures.  \fig{O6_mass_T_bins_SFMS} shows the mass fraction of
circumgalactic OVI in different temperature bins.  At all galaxy
masses, OVI is found primarily in gas in the temperature range
$T=10^{5-6}$ K.  Indeed, the $T=10^{5-6}$ K bin dominates the total
OVI budget to the extent that the total amount of OVI in the CGM
essentially reflects how much $T=10^{5-6}$ K gas is in the halo.  This
implies that collisional ionization is the most important ionization
mechanism, in agreement with earlier works which found that the
high-column OVI absorbers associated with galactic halos are
predominantly collisionally ionized \citep[e.g.,
][]{2011ApJ...731....6S}.  That said, especially in lower mass halos,
a non-negligible contribution to the OVI column arises in cooler,
photo-ionized gas.  This photo-ionized OVI typically occurs within
lower column density absorbers towards the outskirts of the CGM or at
the CGM-IGM interface (see \fig{images}).

\begin{figure}
\includegraphics[width=\newFigurewidth]{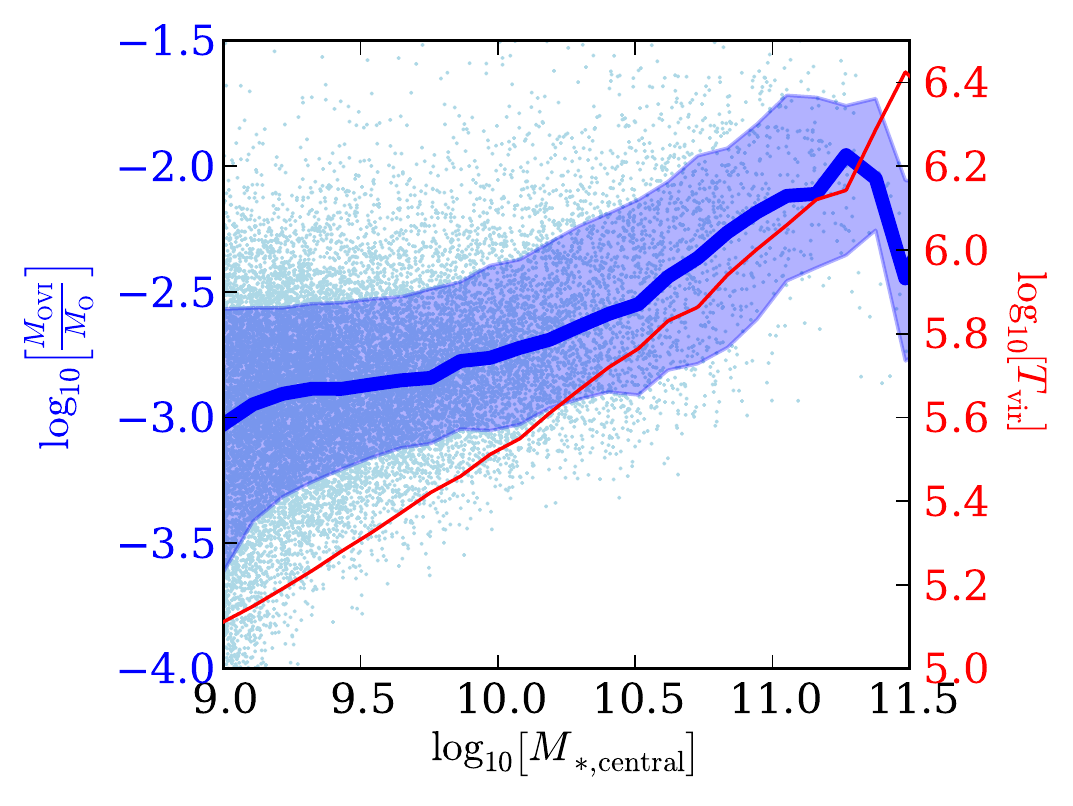}
\label{fig:o6_ion_ratio_SFMS}
\caption{Blue axis: Mass-weighted circumgalactic OVI ionization ratio.
Red axis: Median halo virial temperature.  The ionization fraction
of OVI increases monotonically with galaxy mass as the deeper halo
potential well heats more of the cool gas into the $T=10^{5-6}$ K
range, which OVI predominantly traces (see
\fig{O6_mass_T_bins_SFMS}).  This is true even in cases where
$T_\text{vir}$ exceeds $10^{6}$ K, since the CGM is multiphase.  At
the highest galaxy masses, the ionization fraction drops as the
abundance of $T=10^{5-6}$ K gas declines, with more gas being heated to
$T > 10^6$ K, which is poorly traced by OVI.}
\end{figure}

\fig{T_bins_SFMS} and \fig{O6_mass_T_bins_SFMS} aid interpretation of
how the global circumgalactic OVI ionization fraction scales with
galaxy mass, shown in \fig{o6_ion_ratio_SFMS}.
\fig{o6_ion_ratio_SFMS} shows a monotonic increase in the OVI
ionization fraction between $M_\bigstar=10^{9} M_\odot$ and
$M_\bigstar\sim 10^{11.2} M_\odot$, above which it decreases.  From
\fig{O6_mass_T_bins_SFMS}, this can be understood as an increase in
the abundance of $T=10^{5-6}$ K gas, except for the highest galaxy
masses where there is a downturn.  This is precisely what
\fig{T_bins_SFMS} shows, with the increase in $T=10^{5-6}$ K gas being
matched by a simultaneous decrease in the abundance of $T<10^{5}$ K
gas.  In other words, more of the cool gas in the multiphase halos is
being heated into the OVI ``phase".  Note that since there is always a
significant reservoir of cool gas, regardless of halo mass, this trend
can continue despite halo virial temperatures that are higher than the
peak collisional excitation temperature $T=10^{5.5} K$ of OVI.  This
increase of OVI ionization fraction with galaxy mass finally ends at
the highest masses, where \fig{T_bins_SFMS} shows that the abundance
of $T=10^{5-6}$ K gas decreases as more gas moves to hotter
temperatures.  OVI poorly traces gas at $T>10^6$ K, where OVII, OVIII,
and OIX dominate; at these high galaxy masses there is far more oxygen
in OVII, OVIII, and OIX than in OVI. Unfortunately these higher
species absorb in the soft X-ray range, which limits their
observability.

\begin{figure*}
\includegraphics[width=\wideFigurewidth]{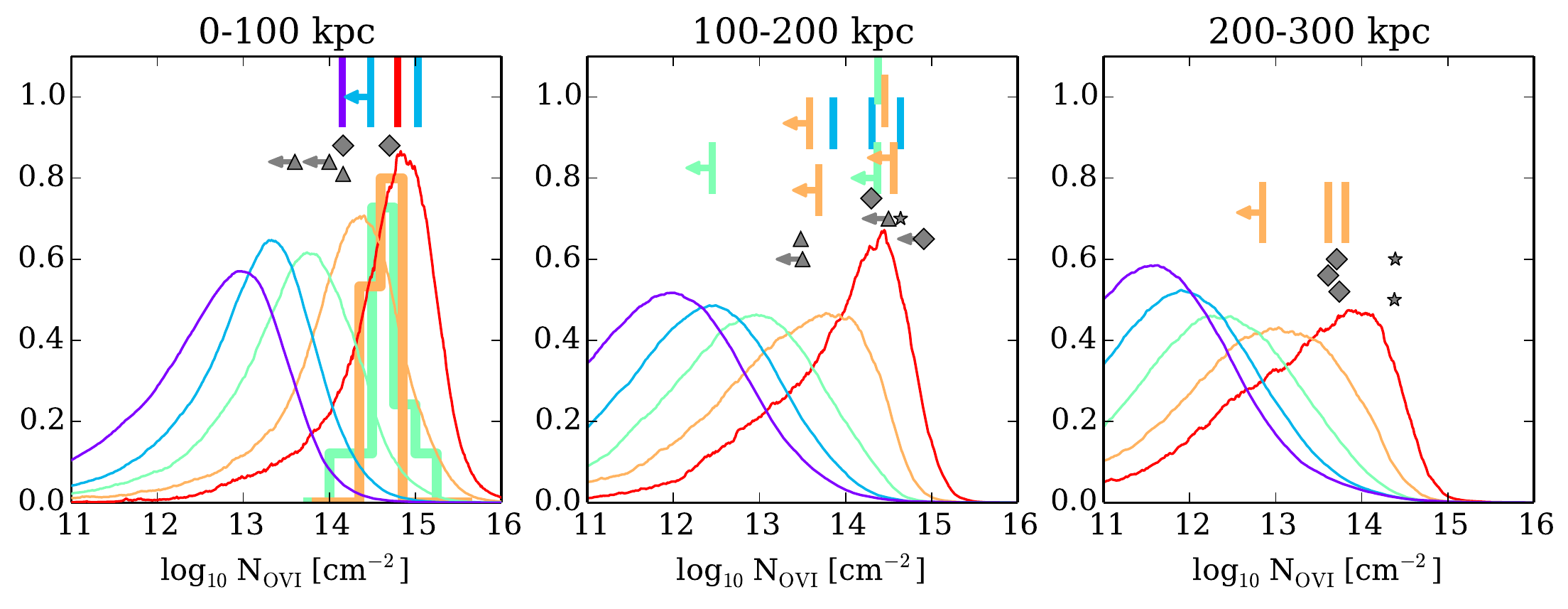}
\includegraphics[width=\subwideFigurewidth]{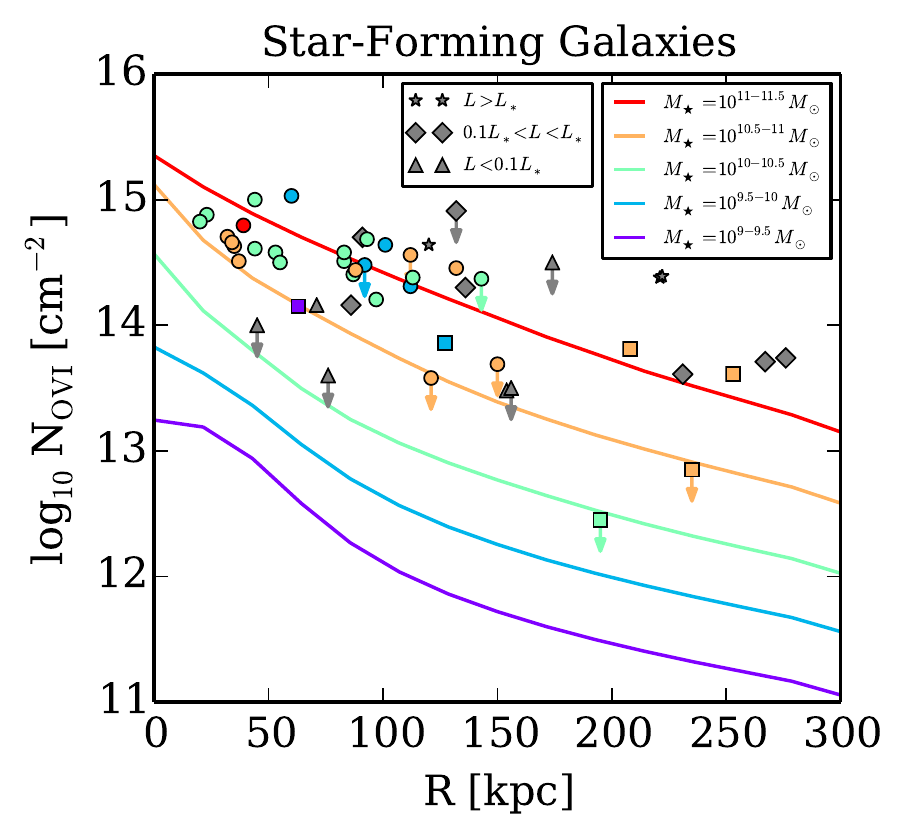}
\label{fig:o6_coldens_SF}
\caption{Model predictions for OVI abundances around star-forming
galaxies of different masses, compared to observations.
\textit{Bottom panel}: OVI abundance as a function of radius, for
various galaxy masses (indicated by different colors).  Curves are
the median profile for star-forming galaxies in given galaxy mass bins,
while observations are shown from \protect\cite{2011Sci...334..948T}
(circles), \protect\cite{2015MNRAS.449.3263J} (squares), and 
\protect\cite{2011ApJ...740...91P} (gray symbols).
\textit{Top panel}: In order to demonstrate the spread about the
median curves in the bottom panel, we show normalized
histograms of all simulated sightlines within three radial ranges.
Colors denote galaxy mass bin, as in the bottom panel.  The thin
curves illustrate the histograms for simulated sightlines, while the
vertical lines indicate individual observed sightlines.  The far left
panel also shows, in thick lines, the histogram of observed
sightlines for the mass ranges $M_\bigstar = 10^{10-10.5} M_\odot$ 
and $M_\bigstar = 10^{10.5-11} M_\odot$ since both of these mass bins
have $\geq 5$ sightlines in this radial interval (note that these
histograms were normalized and then the amplitude was divided by 3 for
visual clarity).  The simulated galaxies show a clear progression in
OVI abundance (more massive galaxies have more OVI), which is
primarily driven by a strong mass-dependence of the mass,
metallicity, and ionization state of the CGM gas.  Conversely, the
observations show a nearly mass-independent OVI profile for all
galaxies above $M \sim 10^9 M_\odot$}
\end{figure*}

\subsubsection{Radial OVI profiles}
\label{sec:SFMS_radial}
In \sect{SFMS_global}, we presented global CGM properties around
star-forming galaxies.  Now we examine the radial profiles of OVI
computed around these galaxies, including both CGM and ISM gas.  The
OVI column densities shown below are derived from all gas within a
line-of-sight velocity cut of $\pm 150$ km/s of the central galaxy.

In \fig{o6_coldens_SF}, we give our model predictions for OVI
abundance profiles around star-forming galaxies with mass $10^9
M_\odot < M_\bigstar < 10^{11.5} M_\odot$.  The model shows a clear
monotonicity where lower mass galaxies have less OVI in their CGM.
This trend does not appear to be reflected in the observations,
although the number of observed sightlines in each galaxy mass bin is
small.  We report the results of log-rank two-sample tests between the
simulations and the observational measurements, computed over radial
bins of width 100 kpc.  The OVI abundance around the two most massive
star-forming galaxies bins ($M_\bigstar = 10^{10.5-11.5} M_\odot$,
shown in orange and red in \fig{o6_coldens_SF}) are consistent with
the observations.  Conversely, the simulated galaxies in the two lower
mass bins ($M_\bigstar = 10^{9.5-10.5} M_\odot$, shown in green and
blue) have a very low probability of being drawn from the same
population as the observations ($P < 0.0001$), in any radial bin.
Finally, few conclusions can be drawn about how the OVI profile around
galaxies in the lowest mass bin considered ($M_\bigstar = 10^{9-9.5}
M_\odot$, shown in purple) compares to observations, since there is
only one point (a log-rank test in the inner 100 kpc gives a
probability $P=0.08$).

The strong galaxy mass dependence of the circumgalactic OVI abundance
can be qualitatively understood from Figures 4, 5, and 9.  First, low
mass galaxies have less circumgalactic gas around them, since they
inhabit less massive halos.  Second, the smaller galaxies also have
less enriched CGM gas.  Finally, even at fixed oxygen mass, the low
mass galaxies have a lower OVI ionization fraction since their virial
temperature is lower.  All of these effects act to decrease the OVI
abundance, resulting in significantly lower OVI abundances around
low-mass star-forming galaxies compared to massive ones.

The apparent lack of such a clear relationship between OVI abundance
and galaxy mass in the observations is puzzling.  One would naively
expect that all of the conditions we find in the simulation would hold
in the real Universe: smaller galaxies should have less CGM gas, this
gas would be relatively less enriched (if the CGM reflects the
mass-metallicity relation of galaxies), and OVI should be a less
abundant ionization state in smaller halos with lower virial
temperatures.  Hence, we would expect smaller galaxies to have less
OVI.  However, this expectation rests on the assumption that OVI gas
traces well-mixed halo gas.  The fact that galaxies across two orders
of magnitude in stellar mass show the same rough OVI profile with
radius (despite significantly different virial radii) implies that OVI
must trace gas which does not reflect the mean physical state of the
global CGM; e.g. gas which is more enriched or denser.  Since
Illustris underpredicts the OVI abundance around low-mass galaxies, we
can infer that physical processes which the simulation cannot capture
may be important (e.g. conduction layers or turbulent mixing layers at
the boundaries between cold and hot phases).  We will return to these
issues in \sect{discussion}.

\begin{figure}
\includegraphics[width=\newFigurewidth]{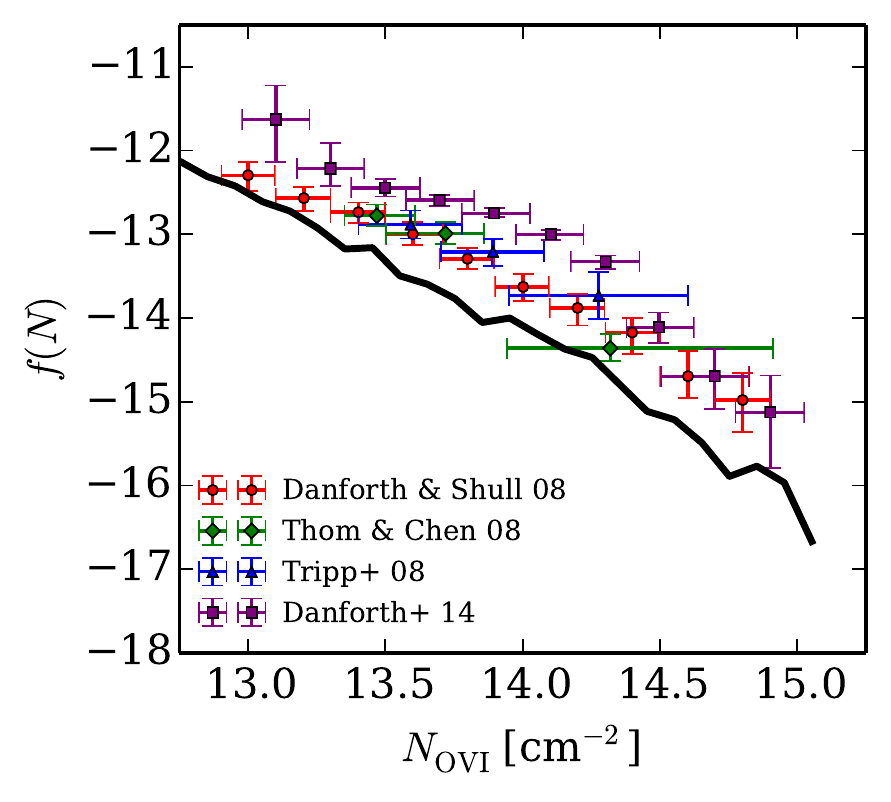}
\label{fig:cddf}
\caption{The column density distribution function (CDDF) of OVI, with the simulation result at $z=0.2$ shown by the black curve.  Observational data comes from \protect\cite{2008ApJ...679..194D}, \protect\cite{Danforth:2014us}, \protect\cite{2008ApJS..179...37T}, \protect\cite{2008ApJS..177...39T}.  The simulation underpredicts the observed CDDF of OVI, reflecting the paucity of OVI around small galaxies. }
\end{figure}

We note that the OVI column density distribution function (CDDF)
generated by the simulation at $z=0.2$ underpredicts observational
constraints (see \fig{cddf}).  Since small galaxies are much more
numerous than large ones, this CDDF underprediction is another
manifestation of the problem that there is too little OVI around small
galaxies.  In \sect{discussion} we explore possible ways for low-mass
star-forming galaxies to generate as much circumgalactic OVI as their
more massive counterparts.

\subsection{Star-forming vs. Quenched Galaxies: $10^{10.5} M_\odot < M_\bigstar < 10^{11.5} M_\odot$}
\label{sec:SFvsquench}

In \sect{SFMS} we investigated the manner in which the mass,
enrichment, and OVI abundance of the CGM depend on stellar mass for
star-forming galaxies.  We now explore the way these CGM properties
depend on the galactic star formation rate, at fixed stellar mass.
Since the COS-Halos survey includes only passive galaxies with mass
$10^{10.5} M_\odot < M_\bigstar < 10^{11.5} M_\odot$, we focus on the
differences between star-forming and passive galaxies in this same
range.  Again, we split the simulated galaxies into categories of
``star-forming" or ``quenched" with a simple cut sSFR $= 10^{-11}$
yr$^{-1}$; the same as used in the COS-Halos survey
\citep[T11;][]{2013ApJS..204...17W}.

In the Illustris simulation, quenched galaxies reside in halos that
are preferentially more massive than those of star-forming galaxies,
at fixed stellar mass (compare, e.g., the right two panels of
\fig{images}).  This implies that the CGM should generally be hotter
around passive galaxies, since the virial temperature is higher.  We
also find that the passive galaxies in our simulation have
systematically more massive central black holes than do the
star-forming galaxies \citep[see also ][]{2015MNRAS.452..575S} of the
same stellar mass.  This is not coincidental, since it is the AGN
feedback that quenches massive galaxies at late times.  Specifically,
V13 showed that the ``radio mode'' of the AGN feedback is most
important for quenching galaxies at low redshift.  As we discuss
below, our implementation of the radio mode physics also has a
significant impact on the circumgalactic gas.

\begin{figure}
\includegraphics[width=\newFigurewidth]{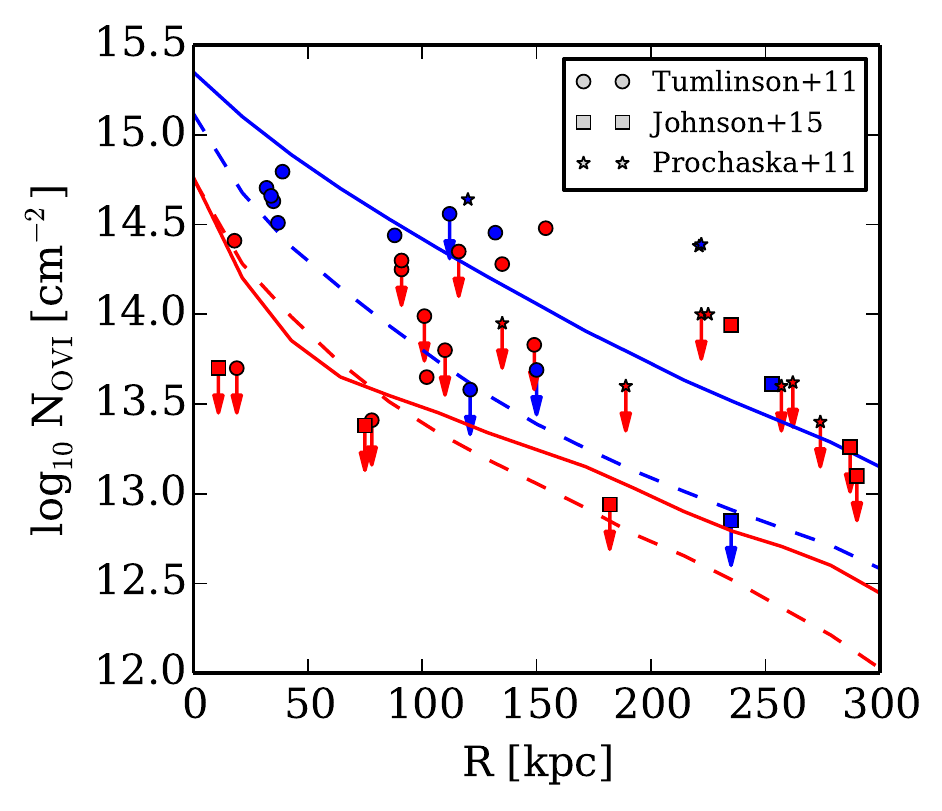}
\label{fig:o6_coldens_SF_passive}
\caption{Model predictions for OVI abundance around star-forming (blue) and
passive galaxies (red) of mass $M_\bigstar = 10^{10.5-11.5}$, compared to observations from
\protect\cite{2011Sci...334..948T} (circles), 
\protect\cite{2015MNRAS.449.3263J} (squares), and \protect\cite{2011ApJ...740...91P} (stars).  
Other than the measurements from \protect\cite{2011ApJ...740...91P}, which are of
 the CGM of galaxies with $L > L_*$, the observational
points are for galaxies of mass $M_\bigstar = 10^{10.5-11.5} M_\odot$.
The curves are median profiles of all simulated galaxies in the
specified mass bin: the solid curve shows $M_\bigstar =
10^{11-11.5} M_\odot$ while the dashed curve is for $M_\bigstar =
10^{10.5-11} M_\odot$.  The observed populations are most different
within the inner 100 kpc.  The simulation reproduces the
underabundance of OVI around passive galaxies compared to their
star-forming counterparts.}
\end{figure}

\fig{o6_coldens_SF_passive} shows the analog of \fig{o6_coldens_SF},
but split instead by passive and star-forming galaxies (red and blue,
respectively), and only for massive galaxies with $10^{10.5} M_\odot <
M_\bigstar < 10^{11.5} M_\odot$.  As in \fig{o6_coldens_SF}, the
observational points are from T11, \cite{2015MNRAS.449.3263J}, and
\cite{2011ApJ...740...91P} (indicated by circles, squares, and stars,
respectively).  The solid curves show the median simulation result for
the mass range $M_\bigstar = 10^{11-11.5} M_\odot$, while the dashed
curves provide the same for $M_\bigstar = 10^{10.5-11} M_\odot$.

When we focus on OVI measurements around star-forming and passive
galaxies in the same stellar mass range ($M_\bigstar = 10^{10.5-11.5}
M_\odot$), as in \fig{o6_coldens_SF_passive}, the two observed
populations appear fairly similar.  Nevertheless, T11 found that a
Kolmogorov-Smirnov test rejects the null hypothesis that they are
drawn from the same parent distribution at $>99\%$ confidence.
Binning into radial bins of 100 kpc and performing 1-D log-rank tests
(which can account for upper limits/non-detections), we find that the
most significant difference between the two populations occurs within
the inner 100 kpc.  For example, for sightlines with impact parameter
$b \leq 100$ kpc, a log-rank test rejects the null hypothesis that
passive and star-forming galaxies are drawn from the same parent
population at $>99\%$ (the difference is especially striking for $b
\leq 50$ kpc, even by eye).  Conversely, for both the impact parameter
bins $100-200$ kpc and $200-300$ kpc, the log-rank test gives
probabilities of $\gtrsim 0.86$, implying that at $r > 100$ kpc the
passive and star-forming galaxy sightlines are consistent with being
drawn from a single parent population.  In other words, the strongest
difference in OVI abundance around star-forming and passive galaxies
appears to occur at low impact parameter ($r \lesssim $50-100 kpc). We
return to the possible significance of this in \sect{SED}.

The simulation reproduces this decrease in the abundance of OVI around
passive galaxies compared to star-forming galaxies.  We will infer the
origin of this bimodality, and why the offset between the OVI
abundance around passive and star-forming galaxies is larger at higher
galaxy mass, by repeating the analysis we performed in \sect{SFMS}.

\begin{figure}
\includegraphics[width=\newFigurewidth]{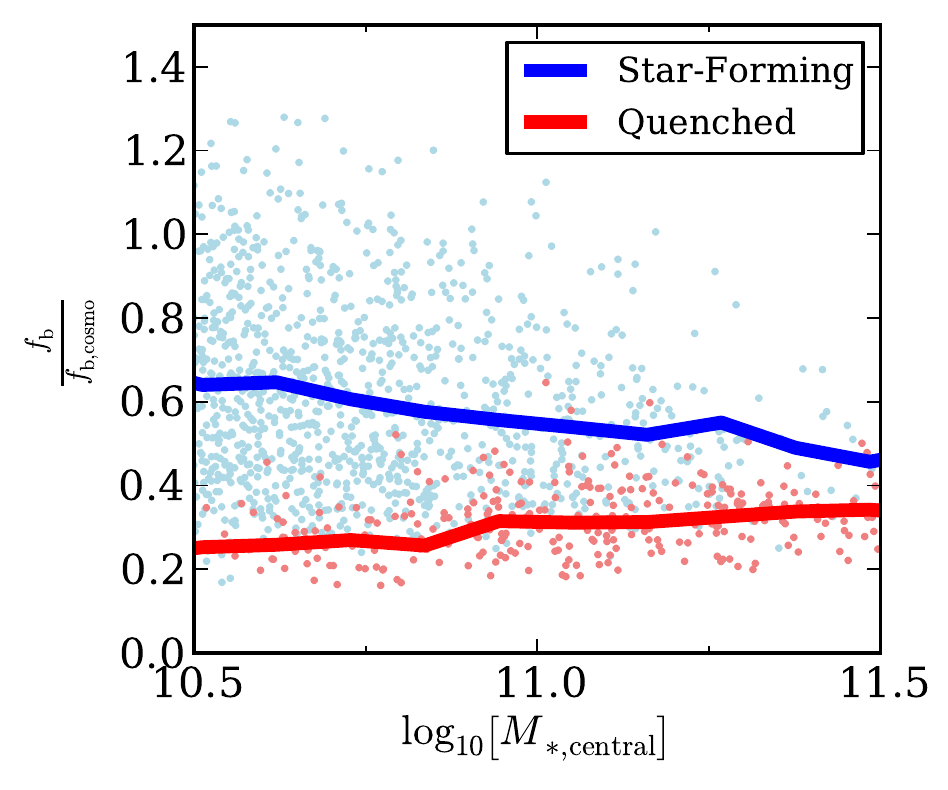}
\label{fig:bary_frac_200_passSF}
\caption{Baryon fraction within R200, as a function of galaxy mass,
split into star-forming and quenched galaxies with the cut of sSFR
$= 10^{-11}$ yr$^{-1}$.  Quenched galaxies have a significantly lower baryon
fraction than star-forming galaxies, due to the action of the
radio-mode component of our AGN feedback model.}
\end{figure}

\fig{bary_frac_200_passSF} shows that the baryon fraction of quenched
galaxies is significantly lower than those that are star-forming.  The
depletion of baryons around passive galaxies in the Illustris
simulation is due to the radio-mode component of our AGN physics
model.  The radio-mode efficiently heats the CGM gas and ultimately
displaces a significant amount of this gas from the inner halo
\citep[see also ][]{2014MNRAS.445..175G}.  Conversely, the galactic
winds, which are present in both star-forming and passive galaxies,
are mostly unable to escape the halo, which allows the star-forming
galaxies to retain high baryon fractions in their halos (see
\fig{baryfrac_SFMS}).  As a result of this radio-mode feedback, we
find that for galaxies with stellar mass $10^{10.5} M_\odot <
M_\bigstar < 10^{11.5} M_\odot$, the median CGM mass of quenched
galaxies is $\sim 0.4$ dex below that of star-forming galaxies,
despite the fact that quenched galaxies inhabit more massive halos
than their star-forming counterparts at fixed stellar mass.

\begin{figure*}
\includegraphics[width=\newFigurewidth]{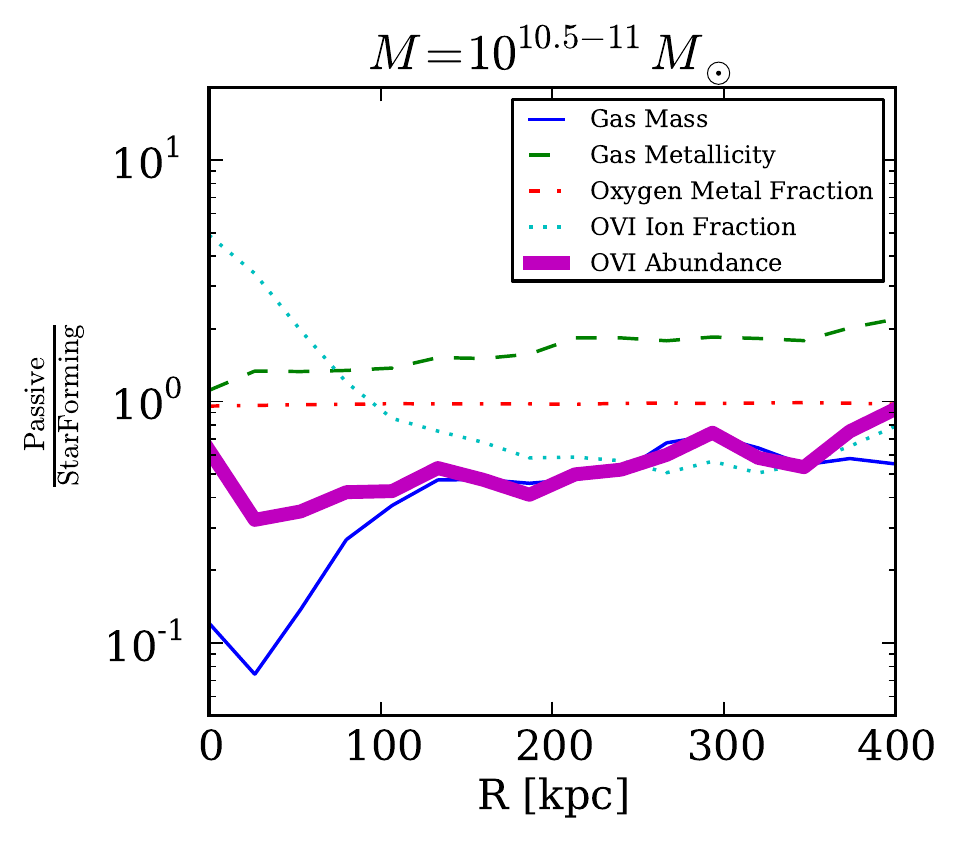}
\includegraphics[width=\newFigurewidth]{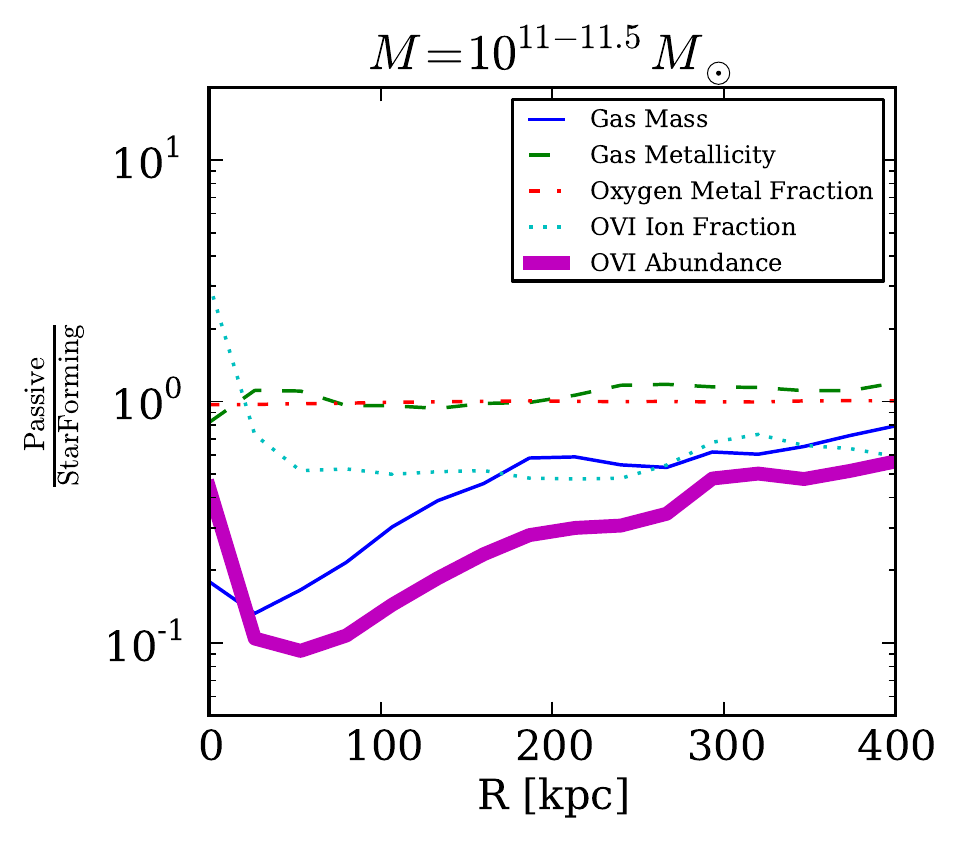}
\label{fig:Mbin_ratios}
\caption{Ratio of radial gas profiles between star-forming and passive
galaxies of mass $M_{\bigstar} =10^{11-11.5} M_\odot$.  The blue
solid curve shows the ratio of the mass profiles, the green dashed
curve shows the ratio of the metallicity profiles, the red
dot-dashed curve shows the ratio of the oxygen-to-metal profiles,
and the cyan dotted curve shows the ratio of the OVI ionization
fraction.  The thick purple curve, which shows the ratio of the
total OVI abundances ($M_\text{OVI,passive}/M_\text{OVI,SF}$), is
simply equal to the product of the four other curves.  The
significant difference in the OVI abundance between passive and
star-forming galaxies is driven by: 1) gas depletion
around the passive galaxies, and 2) the decreased ionization
fraction around the passive galaxies, as the halo moves out of the
OVI phase.}
\end{figure*}

In addition to the difference in CGM mass between star-forming and
passive galaxies, other factors, as shown in \eq{master}, may also
contribute to the OVI abundance offset.  We show the median of the
ratios of each contributing factor, as a function of radius, in
\fig{Mbin_ratios}.  For clarity, the curves denote the following
(where ``SF" denotes star-forming):

\begin{itemize}
\item Blue solid curve: $(M_\text{gas,passive})/(M_\text{gas,SF})$
\item Green dashed curve: $\frac{(M_\text{metals,passive}/M_\text{gas,passive})}{(M_\text{metals,SF}/M_\text{gas,SF})}$
\item Red dot-dashed curve: $\frac{(M_\text{oxygen,passive}/M_\text{metals,passive})}{(M_\text{oxygen,SF}/M_\text{metals,SF})}$
\item Cyan dotted curve: $\frac{(M_\text{OVI,passive}/M_\text{oxygen,passive})}{(M_\text{OVI,SF}/M_\text{oxygen,SF})}$
\item Purple bold curve: $M_\text{OVI,passive}/M_\text{OVI,SF}$
\end{itemize}
Note that the thick purple curve, which shows the total ratio of the
OVI abundance as a function of radius, is simply the product of the
other four.  The two panels of \fig{Mbin_ratios} show the two mass
bins $M_\bigstar = 10^{10.5-11} M_\odot$ on the left and $M_\bigstar =
10^{11-11.5} M_\odot$ on the right.

\fig{Mbin_ratios} allows us to separate out the effects of: 1) CGM
mass, 2) CGM metallicity, 3) oxygen-to-metal ratio, and 4) OVI
ionization fraction.  We see that the mass depletion of the CGM gas,
caused by the AGN radio mode (blue solid curve), is the primary driver
of the difference in the OVI abundance.  The variation in CGM
enrichment between the two galaxy populations (green dashed curve) is
relatively small, within a factor of $\sim 2$, with the passive
galaxies actually having a slightly more enriched CGM on average than
star-forming ones (which can also be seen in the example shown in
\fig{images}).  The difference in the oxygen-to-metal ratio (red
dot-dashed curve) is completely negligible compared to the other
factors.  Finally, the OVI ionization fraction (cyan dotted curve)
varies with radius, such that the passive-galaxy CGM has a higher OVI
ionization fraction than the star-forming-galaxy CGM at small radii,
but has a lower ionization fraction at large distances.

Since passive galaxies reside in more massive halos and also have more
massive black holes, at all radii their circumgalactic gas tends to be
hotter than the corresponding gas around star-forming galaxies.  The
change in the OVI ionization fraction ratio with radius is due to the
multiphase nature of the CGM gas, and where the different phases are
generally located.  There is typically more cool CGM gas close to the
galaxy.  This gas is significantly warmer around passive galaxies than
star-forming ones, leading to a higher fraction of gas in the
temperature range $T\sim 10^{5-6}$ K, which, in turn, boosts the OVI
ionization fraction.  Conversely, gas at larger distances from the
central galaxy is typically virialized, hot gas.  At these distances,
primarily due to AGN heating, the passive-galaxy CGM gas is hot enough
($T > 10^6$ K) to move out of the OVI phase; hence passive galaxies
have a lower effective ionization fraction of OVI than star-forming
ones at large radii.

The boosted ionization fraction at small radii around passive galaxies
competes with the significant gas depletion.  For the lower mass bin
($M_\bigstar = 10^{10.5-11} M_\odot$), the two effects largely cancel,
such that the high ionization fraction at $R \lesssim 100$ kpc
``masks" the depleted CGM, and the total OVI abundance is not
significantly different between the two galaxy populations.  However,
in the higher mass bin ($M_\bigstar = 10^{11-11.5} M_\odot$), all of
the gas is hotter, so the boosting of the ionization fraction at small
radii is less important, except near the galaxy.  Therefore, the
masking due to a boosted OVI ionization fraction is less effective,
and the resulting OVI abundance is lower.  This explains why
\fig{o6_coldens_SF_passive} shows a larger difference in the higher
mass bin (solid curves) than the lower mass one (dashed curves).  In
summary, the simulations predict that the OVI profile around passive
galaxies is primarily driven by a mass-depleted CGM and a
radius-dependent ionization fraction (which is itself a consequence of
a multiphase CGM, with cooler gas at low radii, and hotter gas at
larger radii).

\subsection{The Role of Local Photo-ionization from the Central Galaxy}
\label{sec:SED}

In \sect{SFvsquench} we showed that strong AGN feedback could be a
driver for the difference in OVI abundance around passive and
star-forming galaxies.  However, we know from other observational
probes that the Illustris AGN physics model is too energetic and is
known to expel too much gas from inner halos at masses greater than
$10^{13} M_\odot$.  The differences in \sect{SFvsquench} were largely
driven by the radio mode feedback heating and excavating the inner
halo regions of halo gas.  Fig. 10 of \cite{2014MNRAS.445..175G} shows
that for the most massive objects in Illustris, the inner halo is too
gas-poor.  Furthermore, \fig{o6_coldens_SF_passive} indicates that the
AGN imprints a bimodality in OVI abundance around passive and
star-forming galaxies which extends out to even $\sim 300$ kpc,
whereas the observations, while not able to rule out such a
scenario, suggest that the bimodality may only occur within $b
\lesssim 100$ kpc.  Hence, while AGN feedback may indeed have a
significant impact on properties of the CGM, we also explore a
distinct mechanism to explain the passive/star-forming OVI abundance
bimodality: the effect of the different local galactic radiation
fields on the CGM around the two galaxy populations.

Star-forming galaxies may naturally produce more OVI in their CGM
simply because their stars emit a harder collective SED.  For example,
the population of young stars in star-forming galaxies can generate an
excess of soft x-rays from shock heated gas due to stellar winds and
supernovae \citep{2002A&A...392...19C}.  These soft X-rays can
directly photoionize gas into high ionization states such as OVI
\citep{2010MNRAS.403L..16C,2014MNRAS.437.2882K}, naturally explaining
why the CGM of star-forming galaxies has an excess of OVI over that of
passive galaxies (see also Werk et al. 2015, in prep.).

In order to examine the potential significance of this effect, we
irradiate a single halo of gas extracted from the simulation with a
galaxy SED of varying amplitude to determine how much the local
radiation field influences the OVI abundance.  The galaxy SED we use
is the 5 Myr old stellar population SED of \cite{2002A&A...392...19C}.
This SED includes the blackbody radiation from O and B stars, as well
as soft x-rays from shock heated gas due to stellar winds and
supernovae.  We adopt an escape fraction of 5\% for the lower energy
photons, which have a negligible impact on the OVI ionization fraction
in any case.  For the soft x-rays we assume an escape fraction of
100\% (i.e. that the ISM and CGM is optically thin to photons of these
energies).  In effect, this is an extreme model which gives an upper
limit to the consequences of the radiation field.  The amplitude of
the inferred SED scales with the number of young stars, which we
assume varies linearly with the galaxy star-formation rate.

\begin{figure}
\includegraphics[width=\newFigurewidth]{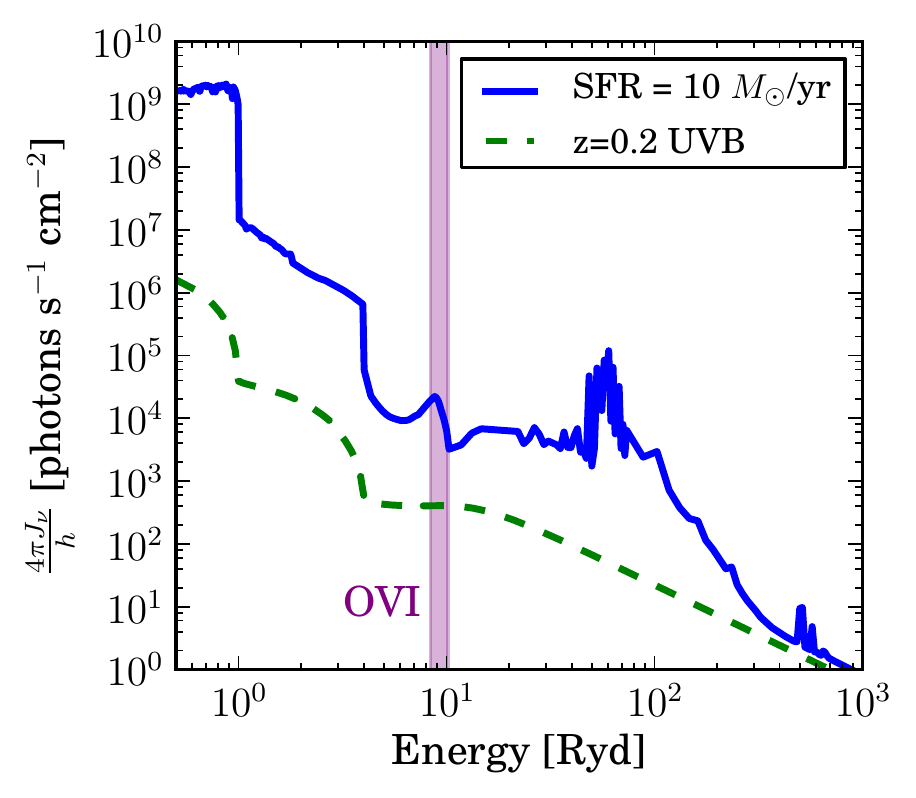}
\label{fig:SED_curves}
\caption{Local galaxy SED (seen here at a distance of 10 kpc), compared 
to the $z=0.2$ extragalactic UVB from \protect\cite{2009ApJ...703.1416F}.
The shaded purple region shows the energy range where OVI is excited.
The small peak of radiation in the galactic SED which aligns with
the excitation range of OVI originates from cooling supernova
remnants.}
\end{figure}

We calculate the effects of the additional, local radiation field on
the simulation in post-processing, again using Cloudy 13.03
\citep{2013RMxAA..49..137F} (see \fig{SED_curves}).  The galaxy SED is
treated as originating from a point-source located at the halo
potential minimum, which is reasonable since the size of the galaxy is
small compared to the scales of interest for the CGM.  This galactic
contribution is then added to the extragalactic UVB from
\cite{2009ApJ...703.1416F} used in the previous sections.

\begin{figure}
\includegraphics[width=\newFigurewidth]{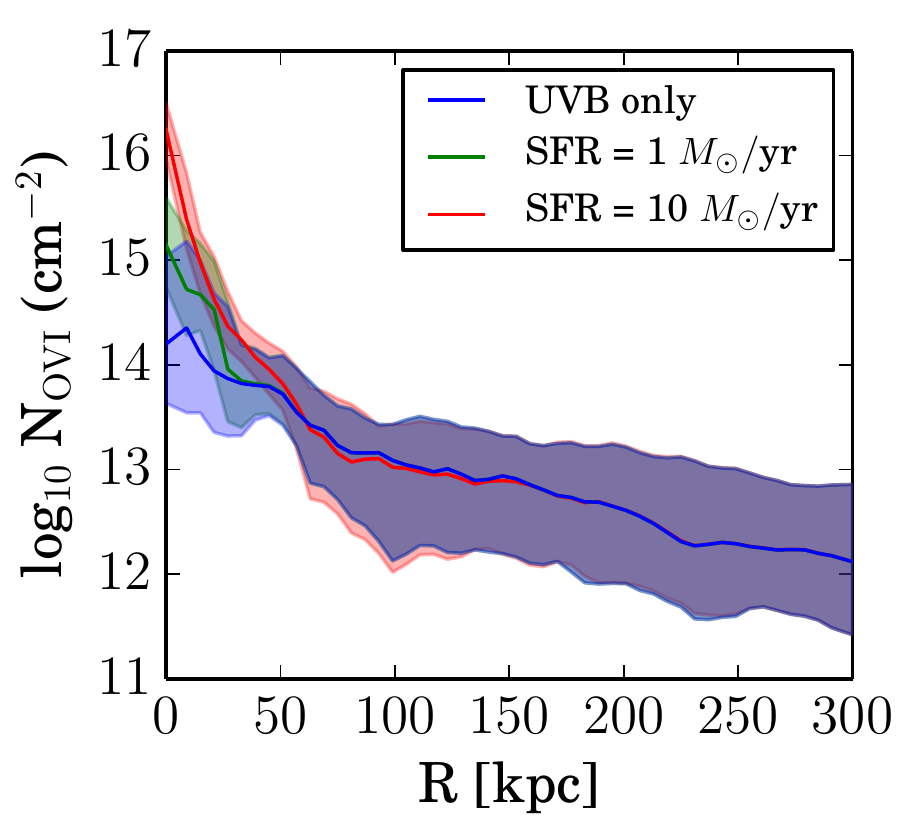}
\label{fig:SED}
\caption{The effect of artificially changing the central galaxy SED
irradiating a typical $10^{12} M_\odot$ halo.  In this model,
increasing the SFR also enhances the soft x-ray component from
cooling of gas shock-heated by supernovae.  This SED
enhancement can produce additional OVI through photoionization
within the inner $\sim 50$ kpc, but has virtually no effect at
larger distances.}
\end{figure}

The object we consider is a single $M\sim 10^{12} M_\odot$ halo at
$z=0.2$ from Illustris, with 3 different SEDs corresponding to SFRs of
zero (extragalactic UVB only), 1 $M_\odot$/yr, and 10 $M_\odot$/yr.
The result is shown in \fig{SED}.  We find that while the local
radiation field has little impact beyond about 50 kpc, there can be
substantial enhancement of the OVI within this radius, due to
photoionization from the galaxy.  When we applied the same simple
approach to all galaxies in the sample, we found a similar result: an
enhancement of OVI around star-forming galaxies within a few tens of
kpc.  This scenario appears consistent with the observations, which
show a stronger bimodality at $r \lesssim 100$ kpc than at larger
distances.  We discuss the difference between this effect and the AGN
in terms of the OVI bimodality in \sect{discussion}.

\section{Discussion}
\label{sec:discussion}

In this work, we have studied the CGM properties and OVI abundance
around galaxies in the Illustris simulation, examining how they vary
with stellar mass and star-formation rate.  We now discuss some of the
implications of our results.

\subsection{Constraints on Galactic Feedback from Star Formation and the Halo Baryon Fraction}
\label{sec:halo_constraint}

We found that the baryon fraction of galaxies is a function of both
stellar mass and star-formation rate.  The baryon fraction decreases
monotonically with increasing stellar mass, and more strongly quenched
galaxies have lower baryon fractions.  These are both
feedback-dependent outcomes.  The baryon fraction is high in
lower-mass galaxies because the wind speed scales with the halo
mass, and the winds are typically not energetic enough to escape the
halo \citep{2015MNRAS.448..895S}.  We have confirmed this result by
inspecting the baryon fraction within smaller simulation volumes which
have varying galactic wind speeds.  Conversely, our AGN model is
energetic enough to launch gas out of the halo, and since more
massive (and more quenched) galaxies typically have larger black
holes, these galaxies lose a more substantial fraction of gas and have
lower baryon fractions.  If we had instead employed stronger or very
stochastic winds which could escape the halo at early times, or an AGN
model which did not significantly affect the halo gas abundance, the
baryon fraction relation with stellar mass could be inverted
\citep[which is the case, for example, in the EAGLE simulation, as
  shown in][]{2015MNRAS.451.1247S}.  In other words, the amplitude and
slope of the baryon fraction versus galaxy mass depends sensitively on
whether feedback is capable of driving gas out of halos
(and, if so, on the galaxy mass scales affected).

The stars and CGM at high and low redshift provide several
complementary constraints on galactic feedback.  In
\cite{2015MNRAS.448..895S}, we found that early enrichment of the
outskirts of massive halos, as observed in, e.g.,
\cite{2014MNRAS.445..794T}, requires metals to escape halos very early
on, which can subsequently expand through the Hubble flow
\citep{2006MNRAS.373.1265O}.  However, \cite{2013MNRAS.436.3031V}
found that with overly energetic winds, the low-redshift cosmic
star-formation rate is too low, since there is insufficient recycling
of halo gas to fuel late-time star formation.  Now we additionally
point out that stronger winds also systematically lower the baryon
fraction of the halo, which is conservatively estimated to be 40\% in
$\sim 10^{12.2} M_\odot$ halos \citep{2014ApJ...792....8W}.

The high measured baryon fractions of \cite{2014ApJ...792....8W} argue
against very strong winds, such as those employed in the EAGLE
simulation \citep{2015MNRAS.451.1247S}.  Unfortunately, the baryon
fraction measurement of \cite{2014ApJ...792....8W} is at
$M_\text{halo} \sim 10^{12.2} M_\odot$, which lies roughly at a
transition point where both galactic winds and AGN are important,
making it difficult to directly place strong constraints on either
feedback channel.  A future baryon fraction measurement for lower halo
masses ($M_\text{halo} \lesssim 10^{11} M_\odot$) would put a stringent
limit on the strength of galactic winds specifically, since AGN
feedback is expected to have little effect in this mass range.

\subsection{The Puzzling Independence of CGM OVI abundance from Galaxy Mass}

One of the aims of our work has been to understand the origin of OVI
absorption around galaxies.  However, we found that the simulation
predicts a much stronger mass-dependence of the OVI column density
profile with radius than is observed.  The apparent lack of
mass-dependence in the observations is surprising, and appears to
require that the OVI abundance be somehow decoupled from the more
global properties of halo gas.  If the OVI abundance actually reflects
the mass, metallicity, and ionization state of the virialized and
well-mixed gas in halos, then it should scale strongly with galaxy
mass, as each of these three quantities increases with galaxy mass.
Since our simulations match observations at higher galaxy masses, the
question seems to be how low-mass galaxies can generate as much OVI as
galaxies which are tens or even a hundred times more massive.

Furthermore, why is it that this seemingly ``universal" OVI profile
for all galaxy masses scales with physical radius in roughly the same
manner, not depending on the virial radius of the galaxies in
question?  We discuss a few leading alternatives as to what may be
driving the OVI excitation around star-forming galaxies, offering various scenarios for the origin of the OVI absorbers: 1) recent winds, 2)
shock-heated WHIM or gas cooling out of the hot CGM, or 3)
photoionization from a soft X-ray feature in the SED of star-forming
galaxies.  Finally, we discuss possible limitations of our simulation arising from our finite mass resolution and the assumption of ionization equilibrium. 

The first scenario, that the OVI originates in highly enriched, recent
outflows, is appealing because it could naturally explain the
bimodality between star-forming and passive galaxies, if active star
formation generates enriched winds.  \cite{Thompson:2015wy} recently
proposed a model in which hot winds, propagating adiabatically, cool
rapidly as they reach the upturn in the cooling function, likely
spending some time in the OVI phase.  However, the cooling time at
this point is short (e.g. the very steep drop in temperature seen in
all models in Fig. 2 of their paper), and hence the OVI phase would
not last long.  It is, therefore, unclear if this mechanism can
explain the high observed column density and covering fraction of
OVI.

As to the reason that the star-forming galaxies across all masses have
nearly the same OVI profile, perhaps the smaller galaxies, although
less enriched than larger ones, are able to launch more gas in winds
due to their shallower potential wells (i.e. they have a higher wind mass-loading).  It may be that the two competing
effects of metallicity and wind mass-loading cancel out, yielding similar profiles at all galaxy masses.
This scenario seems somewhat contrived, however, particularly
considering that the observed ``universal" OVI profile does not appear
to be sensitive to the virial radii of the galaxies in question,
whereas observed wind speeds are weakly correlated with the depth of the potential
\citep{2005ApJ...621..227M, 2009ApJ...692..187W,2014ApJ...794..156R,2015ApJ...811..149C}.   Furthermore, this
effect should be present in the Illustris simulation.  Our theoretical
model features a significantly larger wind mass-loading factor for
smaller galaxies as well as maximal enrichment since the
winds typically remain bound to the halo.  That the simulation
nevertheless predicts a strong galaxy mass scaling casts doubt on the
recent-enrichment scenario.

Another possibility is that the OVI represents the tail end of an even
hotter CGM phase \citep[e.g.,][]{2002ApJ...577..691H}, which may be
gas that has been shock-heated through hierarchical structure
formation \citep{1999ApJ...511..521D, 1999ApJ...519L.109C,2001ApJ...552..473D}.  (Note that typically discussion of the WHIM involves gravitational shocks on up to $\sim$Mpc scales, but here we specifically mean virialized hot gas localized within $R_\text{vir}$.)
This could explain a ``decoupling" between the metallicity of the OVI phase
and that of the aggregate halo gas, since the cooling time is
sensitive to the level of enrichment, so that gas cooling out of this hot phase into OVI
will be preferentially enriched.  Such a scenario may be especially
viable for the more massive galaxies, which contain significant
amounts of gas at temperatures above $10^6$ K that is better traced by
OVII, OVIII, and sometimes OIX.  As it cools, this massive reservoir
of metal-enriched gas can replenish the OVI phase.  Furthermore, if
gas entering this phase continues to cool, it may accrete into the
galaxy and fuel star-formation, partly accounting for the
passive/star-forming OVI bimodality.  However, this mechanism is far less effective in low-mass galaxies, since they have significantly less hot gas (see \fig{T_bins_SFMS}).  Therefore, if this is the
only physical mechanism driving the OVI phase, it is likely that
smaller galaxies would still have a significantly lower OVI abundance
than larger ones, in conflict with the observations.  Furthermore, \cite{2009ApJ...697.1784Y} did not detect OVII and OVIII in stacked spectra of 19 OVI absorbers in six QSO sightlines, implying that at least some OVI absorbers are not associated with a hotter gas phase.

Although collisional ionization can account for a significant amount of OVI at gas temperatures of $\sim$10$^{5.4-5.7}$K (the range where the fraction of oxygen in OVI is maximal) it is possible that a significant amount of OVI is cooler, photoionized gas.  As we showed in \sect{SED}, for example, if there is a significant though as-yet undetected strong excess of soft X-ray photons around 100 eV from star-forming galaxies, they could serve to efficiently photoionize OVI at large distances from the galaxy disk (Cantalupo et al. 2010, Werk et al. 2015, submitted). This scenario is appealing because a local ionizing radiation field from sources within a galaxy could naturally explain the correlation between OVI and galaxy SFR, and thus the dearth of OVI around non-star-forming galaxies.   However, this mechanism alone is insufficient to explain the ``universal" OVI profile for star-forming galaxies, since more massive galaxies, which have more circumgalactic gas and higher metallicity than low-mass galaxies, would likely also have more photo-excited OVI as well.  Second, this effect will be strongest in the most massive galaxies, since their absolute star-formation rates are highest.  Furthermore, as we showed in \fig{SED}, the enhancement of OVI by this effect likely does not extend beyond a few tens of kpc from the galaxy.  Finally, Werk et al. (2015, submitted) show that such models must be highly tuned to explain other aspects of the observations, including universally undetected NV absorption ($\log_{10} N_\text{NV} < 13.5$) and an excess of SiIV absorption compared to lower-ionization state counterparts (e.g. SiII and SiIII).

Could resolution be the issue?  Since the simulation code is adaptive and the CGM consists of relatively low density gas (typically $-5 \lesssim \log_{10} n_\text{CGM} \lesssim -2$), the typical cell size in the CGM is a few proper kpc.  If the OVI is primarily tracing gas which is diffuse and spans the halo (e.g. the warm gas seen in \fig{images}), then our resolution should not be a major concern.  However, it may be that a significant fraction of OVI gas arises in interfaces between cold and hot phases.  These layers may be described by a turbulent mixing or a thermal
conduction layer, depending on the magnetic field (which dictates how
well electrons are able to conduct thermal energy from the hot gas to
cooler regions).  These processes are not resolved in the Illustris
simulation and magnetic fields were not included.  Could these
possibly explain the universal OVI profile?  It is difficult to
imagine that these interfaces would provide the whole solution.  Even
in this case, the total OVI produced likely scales with the
overall abundance and metallicity of the gas, which, again, both
increase with galaxy mass.  The importance of conductive/turbulent
mixing layers between cool and hot regions can only be addressed with
much higher resolution simulations.

Could our assumption of ionization equilibrium be at fault?  We would expect this to be more of an issue for larger galaxies, where AGN fossil zones may be relevant \citep{2013MNRAS.434.1063O}.  However, the OVI profiles of massive galaxies in the simulation are consistent with observations; the
problem is that the simulation underpredicts the abundance of OVI
around low-mass galaxies.  Even if small galaxies do have some periods
of time when they are out of ionization equilibrium, we expect these
periods to be relatively short.  And since we have a large
statistical sample of these smaller galaxies, it seems unlikely that
short periods of non-equilibrium ionization would significantly change
the overall distribution of the column density profiles for this
population.

Most fundamentally, the problem likely stems from the phase structure of the CGM.  Our wind velocity scaling affords a maximal enrichment scenario where the bulk of metals produced by small galaxies are retained in the CGM, a picture that agrees well with the estimates of \cite{2014ApJ...786...54P}.  Furthermore, we expect that the mass and metallicity of the CGM should scale with galaxy mass.  Hence, the only way to recover a roughly uniform OVI profile at all galaxy masses is for smaller galaxies to have a significantly higher global OVI excitation than more massive galaxies.  In other words, the observations seem to require that that smaller galaxies push more of their limited oxygen into OVI, while more massive galaxies hide more of their huge oxygen reservoirs in higher ionization states.  For example, if the real CGM were systematically hotter than in the simulation, it could boost the OVI abundance around small galaxies (which have large reservoirs of cool gas around them), while pushing more of the warm gas in more massive halos over the $T \sim 10^6$ K threshold, out of OVI and into OVII, OVIII, and OIX.

One way to move in this direction would be to simply add more energy into the CGM, for example, by invoking higher energy winds.  However, the wind strength is a parameter which is strongly constrained by the stellar mass function and global star-formation rate, as well as by the halo baryon fraction (see \sect{halo_constraint}).  Another possibility, explored by \cite{2014MNRAS.437.2882K}, is that the local radiation field can couple effectively to the CGM gas and add a photo-heating term to the thermal balance of the CGM, keeping more of the gas heated.  Another, more speculative idea is that cosmic-ray heating could be significant; a process which is not captured in our present simulation.  We leave exploration of these possibilities to future theoretical work.

\subsection{The Passive/Star-forming OVI Bimodality and the Role of AGN in the CGM}
\label{sec:disc_bimodal}
From other studies, we know that the Illustris AGN feedback model is
too extreme.  Fig. 10 of \cite{2014MNRAS.445..175G} shows that for the
most massive systems the inner halo is too gas-poor.  This is caused
by the AGN radio mode implementation, which at late times heats the
inner CGM significantly, driving out significant amounts of gas,
especially around passive galaxies.  X-ray observations, which are
particularly sensitive to the abundance of this inner halo gas ($L_X$ scales roughly as gas density squared),
indicate that the AGN model is too vigorous, especially in massive ellipticals \citep{2015ApJ...804...72B}.

Given this caveat, it is interesting to note that the OVI bimodality we predict between passive and star-forming galaxies is mainly a consequence of
AGN feedback.  The bimodality disappears in smaller simulation volumes without AGN feedback.  Indeed, \cite{2015arXiv150302084F}, who carried out a similar study using a physics model which incorporates a quenching
mechanism that has little effect on the CGM gas, were unable to
reproduce the difference in OVI equivalent width between passive and
star-forming galaxies.  This seems to suggest that while our AGN model
is too extreme, black hole feedback may still play a role in the
reduced OVI abundance around passive galaxies.  In our model, AGN affect the CGM by heating the gas as well as driving gas out of the CGM.  The gas excavation uniformly lowers the OVI abundance, especially at small impact parameters.  The gas heating increases the average OVI ionization fraction at low impact parameter (heating cool gas into the OVI phase), while having the opposite effect at large impact parameter (heating warm virialized gas out of the OVI phase).  We leave a detailed study of the OVI abundance and possible constraints on AGN feedback to future work.

In \sect{SED}, we showed that local photoionization could also help explain the bimodality between passive and star-forming galaxies.  AGN and local photoionization both produce a bimodality, but they act on different galaxy populations.  AGN feedback acts to preferentially suppress the OVI abundance around passive galaxies, by the mechanisms described above.  Conversely, local photo-ionization acts to boost the OVI abundance around star-forming galaxies by photo-ionizing the cool gas located within several tens of kpc from the galaxy.  However, since the largest star-forming galaxies in the simulation already match the observed OVI column density profile (see \fig{o6_coldens_SF}), this implies that the OVI boost from local radiation is small.  Furthermore, the local photo-ionization scenario cannot help boost our low-mass galaxies to the observed ``universal" profile, since the absolute star-formation rates of these small galaxies are generally too low to have a significant impact on circumgalactic scales.

\section{Summary}
\label{sec:summary}

In this work, we have studied the mass, temperature, and enrichment of the circumgalactic medium around galaxies of mass $10^9 M_\odot <M_\bigstar < 10^{11.5} M_\odot$ and with sSFR $< 10^{-9}$ yr$^{-1}$ in the Illustris simulation.  We then use this analysis to explain the abundance of OVI in these environments.  Our main findings are:

1. The baryon fraction of material within the virial radius of these halos decreases with increasing stellar mass.  The high predicted baryon fractions ($\sim 50-100\%$) around star-forming galaxies in halos of mass $M_\text{halo} \sim 10^{12} M_\odot$ matches observational constraints from
\cite{2014ApJ...792....8W}.  While the baryon fraction declines with increasing stellar mass, the overall CGM mass (i.e. the mass of gas within the virial radius that is not associated with galaxies) increases monotonically with stellar mass.  The baryon fraction of the halos hosting quenched galaxies is significantly lower than that of halos hosting star-forming galaxies of the same mass, owing to the more vigorous AGN radio mode which not only suppresses star formation but also drives out significant amounts of gas from the inner CGM.

2. The CGM metallicity reflects the galactic mass-metallicity
relationship, with a similar slope and offset by roughly $-0.3$ dex at all galaxy masses.  However, the CGM metallicity does not vary significantly with
galactic specific star-formation rate.  While the simulation predicts that the CGM of more massive galaxies have a lower oxygen-to-metal ratio than low-mass systems, this difference is negligible compared to the scalings of CGM mass, metallicity, and ionization state with galaxy mass.

3. The CGM has a multiphase character at all galaxy masses.  More
massive galaxies reside in larger, hotter halos and also possess larger central AGN, such that they have a higher mass fraction of gas at higher temperatures.  Nevertheless, even for halos with virial temperatures of $T \sim 10^6$ K, the CGM
mass fraction of gas with $T \lesssim 10^5$ K is $\gtrsim 25\%$.

4. More massive galaxies have higher global OVI ionization fractions
as more of the cool $T \lesssim 10^5$ K gas is heated into the
$T=10^{5-6}$ K range.  This trend is reversed
at the highest galaxy masses ($M_\bigstar \gtrsim 10^{11.2} M_\odot$),
where the abundance of $T=10^{5-6}$ K gas decreases as the abundance
of $>10^6$ K material begins to dominate.

5. Since the CGM mass, metallicity, and OVI ionization fraction all
increase with stellar mass, the simulation predicts that the net OVI
abundance of star-forming galaxies increases significantly with stellar
mass (by $\sim 0.5$ dex increase in OVI column density within 300 kpc of the galaxy position, for each $0.5$ dex increase in stellar mass).  This dependence is not observed. Our model predictions match the observed OVI profiles for star-forming galaxies with $M_\text{galaxy} \gtrsim 10^{10.5} M_\odot$, and underpredict the OVI abundance for lower-mass galaxies by $\gtrsim 0.8$ dex.

6. We reproduce the observed bimodality in the OVI abundance between star-forming
and quenched galaxies for galaxies of mass $10^{10.5} M_\odot < M_\bigstar < 10^{11.5} M_\odot$.  This is caused by AGN feedback, which heats
and drives out gas from the inner CGM.  In the inner regions, cool gas is heated into the OVI phase by the
AGN, while at larger radii, the virialized gas is pushed out of the
OVI phase by AGN heating (as well as by the fact that at fixed stellar
mass, massive quenched galaxies inhabit larger halos than do
star-forming galaxies).  We argue that while our AGN model is known to
be too vigorous, some impact of the AGN on the CGM may be required to
recover the observed bimodality. 

7. The observational data suggests that the largest difference between the passive and star-forming galaxy OVI abundances occurs at $b \lesssim 50$ kpc from the central galaxy.  We argue that, distinct from the effects of AGN, another mechanism influencing the OVI bimodality may be the harder SEDs of
star-forming galaxies compared to passive ones.  Specifically, soft
X-rays originating from processes related to young stars may enhance
photoionization of cool gas near the galaxy, out to distances of tens
of kpc.

We have identified a few promising observational probes which could
further constrain theoretical models of galaxy formation.  First, a
measurement of the baryon fraction at lower halo masses would limit
the strength of galactic winds (and, hence, constrain the
importance of galactic recycling).  Second, more measurements of the OVI column density around low-mass galaxies ($M_\bigstar \lesssim  10^{10} M_\odot$) would tighten constraints on the mass-dependence of the OVI absorption profile.  Third, more measurements of the OVI column density are needed at intermediate distances ($b \sim$ 100-200
kpc) from massive star-forming and passive galaxies, to assess whether the OVI bimodality extends to these impact parameters, or is limited to $b
\lesssim 50$ kpc; such measurements would help determine if the
bimodality is driven primarily by AGN feedback or local galactic
radiation fields.

\section*{Acknowledgements}
JS thanks Dylan Nelson for many helpful discussions and Simeon Bird
for proofreading an earlier draft of this manuscript.  LH acknowledges support from NASA grant NNX12AC67G and NSF grant AST-1312095.


\bibliographystyle{mn2e}
\bibliography{/Users/Josh/Work/Projects/Bibliography/CGM_bib}

\label{lastpage}
\end{document}